\PassOptionsToPackage{capitalize,noabbrev,nameinlink}{cleveref}                                                                                                                                                 
\PassOptionsToPackage{usenames,dvipsnames}{color}
\PassOptionsToPackage{numbers,compress,sort}{natbib}

\documentclass[10pt,conference,letterpaper]{IEEEtran}
\usepackage{times,amsmath,epsfig}

\newcommand{\paperTitle}{Predictive Indexing}
\usepackage{amsmath}
\usepackage{amssymb}
\usepackage{boxedminipage}
\usepackage{xspace}
\usepackage{tabularx}
\usepackage{balance}  
\usepackage[hyphenbreaks]{breakurl}
\usepackage[font={small}]{caption}
\usepackage{graphicx}
\usepackage{subfig}
\usepackage[usenames,table]{xcolor}
\usepackage{cleveref}

\usepackage[numbers]{natbib}
\usepackage[final]{microtype}
\usepackage[scaled]{inconsolata}
\usepackage[T1]{fontenc}
\usepackage{graphicx}
\usepackage{subfig}
\usepackage{soul}
\usepackage{pifont}

\usepackage{booktabs}
\usepackage[end]{algpseudocode}
\usepackage{algorithm}

\newcommand{\mlbegin}{\shortstack\bgroup}
\newcommand{\mlend}{\egroup}

\newcommand{\squishitemize}{
 \begin{list}{$\bullet$}
  { \setlength{\itemsep}{0pt}
     \setlength{\parsep}{3pt}
     \setlength{\topsep}{3pt}
     \setlength{\partopsep}{0pt}
     \setlength{\leftmargin}{1.95em}
     \setlength{\labelwidth}{1.5em}
     \setlength{\labelsep}{0.5em} } }

\newcounter{Lcount}
\newcommand{\squishlist}{
    \begin{list}{\arabic{Lcount}. }
   { \usecounter{Lcount}
        \setlength{\itemsep}{0pt}
        \setlength{\parsep}{3pt}
        \setlength{\topsep}{3pt}
        \setlength{\partopsep}{0pt}
        \setlength{\leftmargin}{2em}
        \setlength{\labelwidth}{1.5em}
        \setlength{\labelsep}{0.5em} } }

\newcommand{\squishend}{\end{list}}

\definecolor{todo-color}{rgb}{1,0,0}

\definecolor{comment-color}{rgb}{0.25,0.25,0.25}
\newcommand{\codeComment}[1]{\textnormal{\color{comment-color}{\textit{\textbf{\# #1}}}}\unskip}

\newcommand{\benchTuner}{TUNER\xspace}

\newcommand{\narrow}{narrow\xspace}
\newcommand{\wide}{wide\xspace}

\newcommand{\pred}{predictive\xspace}
\newcommand{\Pred}{Predictive\xspace}

\newcommand{\peloton}{DBMS-X\xspace}

\newcommand{\full}{\texttt{FULL}\xspace}
\newcommand{\vbp}{\texttt{VBP}\xspace}
\newcommand{\vap}{\texttt{VAP}\xspace}

\newcommand{\sysPage}{page\xspace}

\newcommand{\tunerF}{{\sffamily\small FAST}\xspace}
\newcommand{\tunerM}{{\sffamily\small MOD}\xspace}
\newcommand{\tunerS}{{\sffamily\small SLOW}\xspace}
\newcommand{\tunerD}{{\sffamily\small DIS}\xspace}

\begin{document}

\newcommand{\mail}[1]{\href{mailto:#1}{#1}}

\title{\paperTitle}

\author{\IEEEauthorblockN{
Joy Arulraj\IEEEauthorrefmark{1}
Ran Xian\IEEEauthorrefmark{2} 
Lin Ma\IEEEauthorrefmark{3}, 
Andrew Pavlo\IEEEauthorrefmark{4}
}
\IEEEauthorblockA{Carnegie Mellon University\\
Email: \IEEEauthorrefmark{1}jarulraj@cs.cmu.edu,
\IEEEauthorrefmark{2}rxian@cs.cmu.edu,
\IEEEauthorrefmark{3}lin.ma@cs.cmu.edu,
\IEEEauthorrefmark{4}pavlo@cs.cmu.edu
}}

\maketitle

\begin{abstract}
There has been considerable research on automated index tuning in database
management systems (DBMSs).
But the majority of these solutions tune the index configuration by 
retrospectively making computationally expensive physical design changes
all at once.
Such changes degrade the DBMS's performance during the process, 
and have reduced utility during subsequent query processing due to the delay
between a workload shift and the associated change.
A better approach is to generate small changes that tune the physical design
over time, forecast the utility of these changes, and apply them ahead of time
to maximize their impact.

This paper presents \pred indexing that continuously improves a
database's physical design using lightweight physical design changes.
It uses a machine learning model to forecast the utility of these changes, 
and continuously refines the index configuration of the database to handle
evolving workloads.
We introduce a lightweight hybrid scan operator with which a DBMS can make
use of partially-built indexes for query processing.
Our evaluation shows that \pred indexing improves the throughput of a DBMS by
3.5--5.2$\times$ compared to other state-of-the-art indexing
approaches. We demonstrate that \pred indexing works seamlessly with other
lightweight automated physical design tuning methods.
\end{abstract}

\section{Introduction}
\label{sec:introduction}

The performance of modern data-driven applications is often constrained by that
of the underlying DBMS. The physical design of the database plays a dominant
role in determining the system's performance. 
Thus, it is important to tune it based on the application's query workload. 
One key component of the physical design problem is to determine a set of
indexes that balances the trade-off of accelerating query execution 
and reducing index maintenance overhead.
Database administrators (DBAs) address this problem using \textit{index
advisors} that are offered by most database
vendors~\cite{chaudhuri97,zilio04}.
These tools recommend indexes to build based on the current index configuration
of the database and the query workload.

Index advisors require a DBA to provide a representative workload collected over
some period of time (e.g., several weeks).
But in modern hybrid transaction-analytical processing (HTAP) workloads, 
it is unlikely that an index is globally useful over the entire
workload~\cite{bruno07,petraki15}.
DBAs must, therefore, continuously monitor and re-tune the index configuration
to ensure that indexes that are appropriate for the current query
workload are available.
Such an \textit{offline} index tuning process involves performing large
computationally expensive index configuration changes.
DBAs are required to carefully schedule these index building operations
during non-peak hours or offline maintenance breaks to minimize their impact 
on the application. With these offline approaches, the DBMS cannot react in
time to workload changes.

Several indexing approaches have been proposed to address the limitations of
offline index advisors, such as 
online indexing~\cite{bruno07,schnaitter06}, 
adaptive indexing~\cite{idreos07}, 
self-managing indexing~\cite{voigt13}, and 
holistic indexing~\cite{petraki15}.
\textit{Online} indexing obviates the need for manually scheduling index
configuration changes by automatically monitoring the workload and
refining the index configuration.
It still, however, involves computationally expensive changes, 
such as immediately building an entire index, 
which degrade the performance of the DBMS while they are being applied.
\textit{Adaptive} and \textit{self-managing} indexing schemes instead
advocate an incremental approach towards index tuning, wherein the index
configuration of the DBMS is evolved using smaller steps.
They build indexes partially and incrementally during query processing, 
and amortize the overhead of constructing an ad-hoc index across several
queries. The main limitation of these schemes is that they only refine the
indexes during query processing, and do not leverage idle system resources when
the DBMS is not processing any queries.
\textit{Holistic indexing} overcomes this limitation by enabling the DBMS to use
idle system resources for optimistically evolving the index configuration.

All of these indexing approaches examine the recent query work- load in
hindsight to determine the set of indexes that must be refined next. This
retrospective approach towards index tuning increases the delay between a
workload shift and the associated physical design change, thereby reducing the
utility of the indexes.     
Another limitation is that these approaches can still degrade the performance of
the DBMS while applying the index configuration changes.
For example, even with holistic indexing, the index tuner might start populating
a substantial part of an index while processing a query to materialize all the
index entries matching the query's predicate. This increases the query's
latency. Such latency spikes may prevent the DBMS from honoring latency
service-level agreements.

In this paper, we present \textit{\pred indexing} that uses a machine learning (ML)
model to predict the utility of indexes in future, and adapts the index
configuration of the database \textit{ahead of time} to increase the utility of
indexes. We propose a simple lightweight \textit{value-agnostic hybrid scan}
operator that allows the DBMS to use partially-built indexes without incurring
latency spikes. A summary of the differences between \pred indexing and other
state-of-the-art indexing approaches is presented in \cref{tab:tuning}.

\begin{table}[t!]
    \centering
    {\small {

\newcolumntype{b}{X}
\newcolumntype{Y}{>{\centering\arraybackslash}X}
\newcolumntype{Z}{>{\centering}p{2cm}}
\begin{tabularx}{\columnwidth}{ZYYYYY}
\textbf{Indexing Approach} & \textbf{Workload Type} & \textbf{Index Type}
& \textbf{Always On} & \textbf{Decision Logic} &
\textbf{Hybrid Scan} 
\\

\midrule

Offline~\cite{finkelstein88,chaudhuri97} & Static &
Full & $\times$ & $\times$ & $\times$ \\

\midrule

Online~\cite{bruno07,schnaitter06} & Dynamic & Full & $\checkmark$ &
Retrospective & $\times$ \\

\midrule

Adaptive~\cite{idreos07},\\ Self-managing~\cite{voigt13} & Dynamic &
Partial & $\times$ & Immediate & Value-based \\

\midrule

Holistic~\cite{petraki15} & Dynamic & Partial & $\checkmark$ & Immediate &
Value-based \\

\midrule

Predictive & Dynamic & Partial & $\checkmark$ & Predictive & Value-agnostic \\

\end{tabularx}
}
}
    \caption{
        \textbf{Comparison of Automated Indexing Approaches} --
        \normalfont{Qualitative differences among the different indexing schemes
        with respect to the types of supported workloads and indexes
        built, the ability to always evolve the index configuration, 
        and the types of decision logic and hybrid scan employed.}
    }
    \label{tab:tuning}
\end{table}

To evaluate our approach, we implemented the \pred index tuner and 
the hybrid scan operator in \peloton, an in-memory HTAP DBMS that is
designed for autonomous operation~\cite{peloton}.
We examine the impact of \pred indexing on the DBMS's performance and its
ability to handle evolving workloads.
Our results show that \pred indexing improves \peloton{'s} throughput
by 3.5--5.2$\times$ compared to other state-of-the-art indexing 
approaches.
In summary, this paper makes the following contributions:

\squishitemize
    \item
    We present a \pred indexing approach that uses reinforcement learning 
    to forecast the utility of the index configuration and tunes it ahead of
    time.
    We propose a lightweight value-agnostic hybrid scan operator with
    which the DBMS can make use of partially-built indexes. 

    \item
    We implement an end-to-end index tuner that decides both \underline{when}
    and \underline{how} to adapt the index configuration.
    We integrate different indexing approaches in the same DBMS, 
    including (1) online~\cite{bruno07,schnaitter06}, 
    (2) adaptive~\cite{idreos07}, 
    (3) self-managing~\cite{voigt13}, 
    (4) holistic~\cite{petraki15}, 
    and (5) predictive.
    We also develop a new open-source benchmarking suite to perform a detailed
    comparison of these indexing approaches~\cite{peloton}.

    \item
    We demonstrate that \pred indexing works seamlessly with other automated
    physical design tuners, and is a step forward towards DBMSs designed for 
    autonomous operation. By working in tandem, the \pred index tuner and
    the storage layout tuner solve two key components of the physical design
    problem.
\squishend

The remainder of this paper is structured as follows. 
We first discuss the benefits of predictive indexing, and then highlight the
performance impact of the hybrid scan operator on HTAP workloads in
\cref{sec:motivation}.
Next, in \cref{sec:hybrid}, we describe the design of the hybrid scan
operator. We then present the predictive ML model that forms the
underlying decision logic of the tuner in \cref{sec:model}, 
followed by our index tuning benchmark in \cref{sec:benchmark}.
We then present our experimental evaluation in \cref{sec:exps}. 
We conclude with a discussion of related work in \cref{sec:related::work}.

\section{Motivation}
\label{sec:motivation}

An important aspect of modern database applications is that their query patterns evolve 
over time. These changes reflect the hourly, daily, or weekly
processing cycles of an organization. For instance, consider a stock exchange application. 
During regular trading hours, it generates a write-intensive online
transaction processing (OLTP) workload. After the trading hours are over, it
mostly executes read-intensive online analytical processing (OLAP) queries that
examine the data collected during the day. At the end of every week, it runs more
complicated analytics to generate reports and summaries.

Several indexing approaches have been proposed to allow the DBMS to handle such
dynamic workloads by automatically adapting the index configuration during
different workload
\textit{phases}~\cite{petraki15,voigt13,idreos07,bruno07,schnaitter06}. These
approaches have two main limitations.
First, they can degrade the performance of the DBMS while applying a physical
design change. For example, the tuner might start populating a substantial
part of an ad-hoc index while processing a query, thereby increasing the query's 
latency.
Second, the utility of the physical design change is dampened due to the delay
between a workload shift and the associated change. 
We refer to this delay as the tuner's \textit{reaction time}.
This reaction time is comprised of two components: (1) the \textit{detection time} is how long 
it takes to detect a workload shift and pick a physical design change, and
(2) the \textit{adaptation time} is how long it takes to apply the change and leverage it 
during query processing. Longer reaction times decrease the utility of indexes
that are dynamically built to accelerate query processing.

\Pred indexing shrinks the tuner's \underline{detection time} by employing a ML
model to forecast the utility of the changes so that they can be applied ahead of time. 
It incrementally builds indexes using lightweight changes to
prevent latency spikes. The DBMS employs a hybrid scan operator to leverage
these partially-built indexes during query processing, and thereby shrinks the
\underline{adaptation time}.
\\ \vspace{-0.1in}

\textbf{Target Workloads:}
\Pred indexing accelerates production HTAP workloads containing both
\textit{recurring} and \textit{ad-hoc} queries.
As the workload evolves over time with recurring queries being added and
removed, the \pred index tuner learns to forecast the utility of the index
configuration, and adapts it ahead of time using lightweight changes.
This shrinks the DBMS's reaction time on evolving HTAP workloads, 
and obviates the need for periodic manual tuning.
We next discuss how \pred indexing differs from other state-of-the-art 
approaches, and then highlight the performance impact of hybrid scan on HTAP
workloads in~\cref{sec:motivation-hybrid}.

\subsection{Predictive Decision Logic}
\label{sec:motivation-predictive}

Every indexing approach is driven by a \textit{decision logic} (DL) 
that determines the type of physical design changes made, and when 
these changes are applied.
Online indexing schemes employ a \textit{retrospective DL} 
that examines the history of the last $k$ queries executed to
determine the set of indexes that could have helped accelerate processing
those queries~\cite{bruno07,agrawal06,schnaitter09}. 
The main limitation of retrospective DL is that it increases the
detection time of the tuner as it needs to examine a longer window of queries.
Longer reaction times of retrospective DL decrease the utility of the changes.

In contrast, adaptive, self-managing, and holistic approaches adopt an 
\textit{immediate DL} ($k$ = 1)~\cite{idreos07,voigt13, petraki15}. 
With immediate DL, the tuner only examines the most recent query while picking
an index for construction, and thus has a shorter detection time. 
But it cannot guard against noisy workloads with one-off queries 
as it immediately starts building indexes to accelerate them. 
With a retrospective DL, the tuner would not immediately construct these indexes
as it examines a longer window of queries to estimate their utility before
building them.

\textit{Predictive DL} addresses the limitations of these two approaches.
It uses a statistical estimator that captures workload trend and seasonality to
forecast the utility of indexes.
For example, the predictive index tuner can detect that a workload shift occurs
every day at 8am, and then build a candidate index ahead of time at
7am, so that the index is immediately available for query processing after the
workload shift. Thus, this forecasting technique reduces the tuner's detection
time. Like retrospective DL, it examines a longer window of queries before
picking a physical design change. This increases its detection time when it 
encounters a one-off query for the first time, but once it captures the query's
trend, it detects it ahead-of-time. In this manner, predictive DL can both
reduce the tuner's detection time, and guard against noisy workloads with
one-off queries. We next present how our hybrid scan operator helps
shrink the tuner's adaptation time.

\subsection{Hybrid Scan}
\label{sec:motivation-hybrid}

\begin{figure}
    \centering
    \includegraphics[width=0.9\columnwidth]{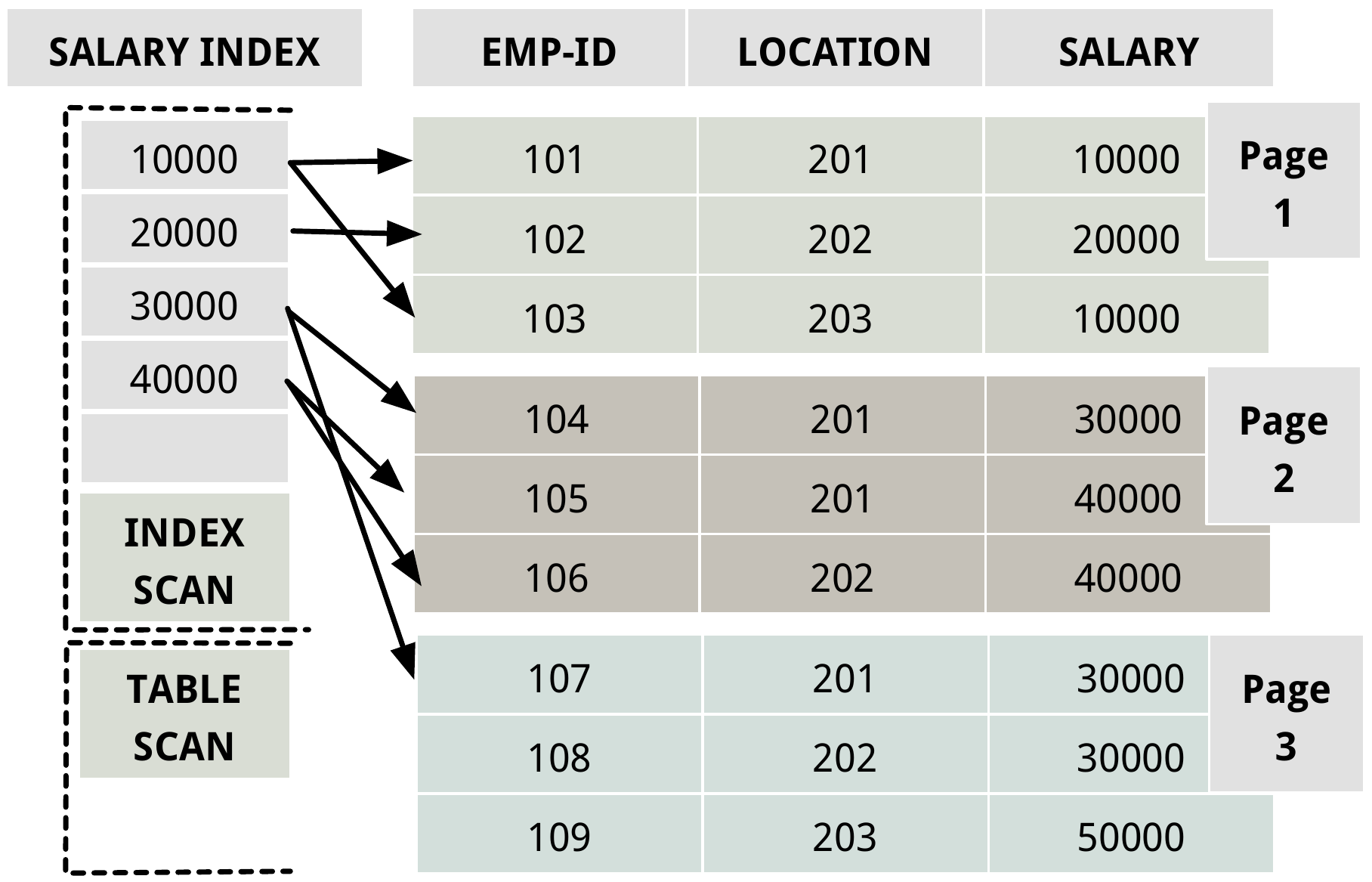}
    \caption{
        \textbf{Hybrid Scan} --
        \normalfont{Value-agnostic hybrid scan over the \texttt{EMPLOYEE} table and the
        partially-built ad-hoc index on its \texttt{SALARY} attribute.}
    }
    \label{fig:incremental-index-tuning}
\end{figure}

The DBMS can make use of ad-hoc indexes constructed by an index tuner in three
ways. We illustrate the differences between these approaches using an example. 
Consider the \texttt{EMPLOYEE} table shown in
\cref{fig:incremental-index-tuning}; assume that it is not indexed.
Suppose the application executes multiple analytical queries that compute the
total salary of all employees whose salary falls within a range:
\\ \vspace{-0.1in}

\begin{center}
\begin{boxedminipage}{0.9\columnwidth}
\texttt{ \textbf{SELECT} \textbf{SUM}(salary) \textbf{FROM} EMPLOYEE \\
\hspace*{4pt} \textbf{WHERE} salary > \ensuremath{t_1} \textbf{AND}
salary < $t_{2}$} 
\label{ex:analytical}
\end{boxedminipage}
\end{center}    

In this scenario, the tuner would observe that the predicate in these queries
repeatedly accesses the \texttt{salary} attribute, and build an index to
accelerate these queries.
Among the three approaches for using this index, the first one is to
start using it only after it is fully populated. 
This is referred to as the \textit{full} scheme (\full), and is employed by
online indexing~\cite{valentin00,chaudhuri98b,sattler2003}.
The main limitation of this approach is that it increases the adaptation time,
thus shrinking the utility of the index during query processing.
A better way is to use the index even when the DBMS has not finished populating it.

Holistic, adaptive, and self-managing indexing approaches adopt such a
\textit{value-based partial} scheme (\vbp)~\cite{idreos07,voigt13,
petraki15}.
With \vbp, the tuner adds index entries based on the values 
present in the predicate of the queries.
For the \texttt{EMPLOYEE} table query, the tuner will add index entries 
for employees whose \texttt{salary} $\in (t_1, t_2)$.
The DBMS can, then, efficiently handle subsequent queries accessing the same
sub-domain of the \texttt{salary} attribute using the index. However, it can do
so only after the sub-domain is completely indexed.
Another limitation of \vbp is that it can still degrade the DBMS's performance
while building the index. When the number of employees whose
\texttt{salary} $\in (t_1, t_2)$ is large, the tuner immediately adds index
entries for those employees while processing the query, thus causing latency
spikes.
Furthermore, the tuner must maintain an additional data structure (e.g., a covering
tree~\cite{voigt13}) for each index to keep track of the indexed sub-domains, 
so that the DBMS can determine if it can process the query using the index.

We propose a \textit{value-agnostic partial} scheme (\vap), where the tuner 
only adds entries for tuples contained in a fixed number of
\sysPage{s} at a time during each tuning cycle, irrespective of the value of the
\texttt{salary} attribute in these tuples.
With \vap, the DBMS can use an index to accelerate query processing without
having to immediately populate the $(t_1, t_2)$ sub-domain. 
This design ensures that the index construction overhead is independent of
the value distribution of the \texttt{salary} attribute. 
The DBMS reads the partially-built index for the already
indexed \sysPage{s}, and falls back to a table scan for the other \sysPage{s}.
With \vap, the tuner does not need to maintain any additional data structures. 
Decoupling index construction from query processing ensures that the tuner
only performs lightweight physical design changes, thereby preventing latency
spikes. We defer a detailed description of this scheme to \cref{sec:hybrid}. 

We present a motivating experiment to illustrate the performance implications of
employing these three different schemes. We load 10m tuples into the
\texttt{EMPLOYEE} table whose \texttt{salary} attribute is an integer value from
a Zipf distribution in the range $[$1,1m$]$~\cite{zipf}.
We configure the input parameters $t_{1}$ and $t_{2}$ in the query's
predicate such that it selects 1\% of the employees. We execute 5000 queries of
the same query type with different input parameters.

\begin{figure}[t!]
    \centering
    \fbox{\includegraphics[width=0.45\textwidth, trim=2.25cm 0.5cm 0cm 0.15cm]
                        {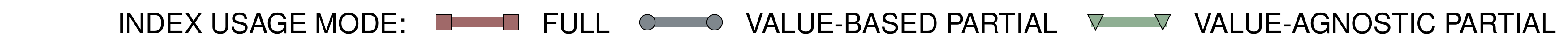}}
    \\[-0.3ex] 
    \includegraphics[width=0.48\textwidth]{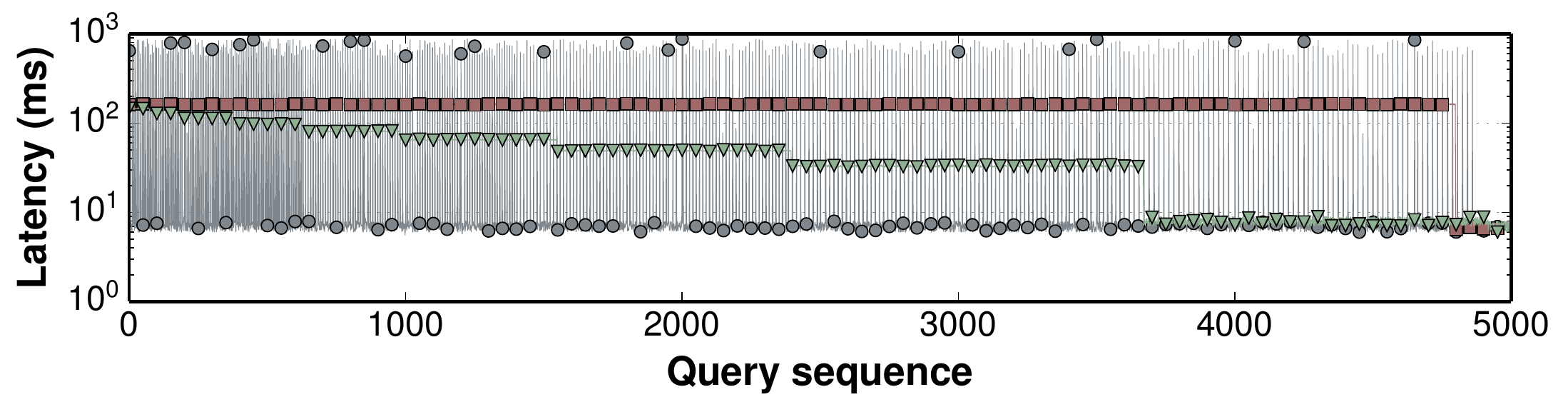}
    \caption{
        \textbf{Ad-hoc Index Usage Schemes} --
        \normalfont{This time series graph illustrates the benefits of using a
        partially-built index during query processing.
        While the query latency follows a bimodal distribution
        with \vbp, it gradually drops over time with \vap.}
    }
    \label{fig:index-usage-modes}
\end{figure}

The results are shown in the time series graph in \cref{fig:index-usage-modes}.
We observe that with \full, the query latency drops sharply after the tuner
completely builds the index. Till then, the DBMS scans the table during
query processing. With \vbp, the query latency follows a bimodal
distribution. The latency matches that of an index scan when the query accesses
an already indexed sub-domain, but then degenerates to a table scan when
it retrieves tuples from a sub-domain that has not been indexed yet.
In the latter case, the tuner immediately populates entries associated with the
sub-domain in the index while processing the query, thus causing latency spikes.

With \vap, the query latency gradually drops over time. This is because the
tuner incrementally populates the index over several tuning cycles, and the
hybrid scan operator increasingly leverages it to skip scanning parts of the
table. A key observation from the time series graph is that \vap prevents
latency spikes.
After the tuner fully populates the index, the DBMS executes the
query 10.1$\times$ faster than a table-scan. 
The cumulative time taken by the DBMS  to execute this workload with \vap is
1.6$\times$ and 3.2$\times$ shorter than that taken with the \vbp and \full
schemes, respectively. The tuner periodically pauses the tuning process to
control the index construction overhead. The impact of the partial schemes
(\vap and \vbp) is more pronounced during such pauses, as they enable the DBMS
to leverage the partially-built indexes.

The main issue with using a partially-built index is that since the tuner
is dynamically populating the index, unless care is taken, 
the hybrid scan operator will return the same matching tuple twice.
Moreover, it may not be able to find a matching tuple if that tuple has
not yet been inserted into the index. 
In the next section, we describe how to solve these problems with our hybrid
scan operator. We then present the \pred index tuner in \cref{sec:model}.

\section{Hybrid Scan}
\label{sec:hybrid}

The hybrid scan operator is a combination of an index and table scan.
We illustrate it using the query on the \texttt{EMPLOYEE} table that 
contains nine tuples stored across three \sysPage{s}.
\cref{fig:incremental-index-tuning} shows the point when the tuner
has finished indexing the tuples contained in the first two \sysPage{s}, and is
in the process of indexing the  third \sysPage. It has indexed only the first
tuple in the third \sysPage, and suppose that this tuple satisfies the query's
predicate.
While processing this query, the hybrid scan reads the partially-built
index to cover the first two \sysPage{s} and then resorts to a table scan over
the third \sysPage. It then returns the tuples retrieved using these two
operations.

This design ensures that the DBMS finds all the tuples satisfying the predicate because the operator 
covers every \sysPage at least once when scanning the index and table.
It can, however, still return the same matching tuple twice, if it is retrieved 
by both operations. 
We next describe how hybrid scan removes 
such duplicate tuples.

\begin{figure}
    \centering
    \subfloat[$\rho_{m}=\rho_{i}$]{
        \includegraphics[width=0.3\columnwidth]
                        {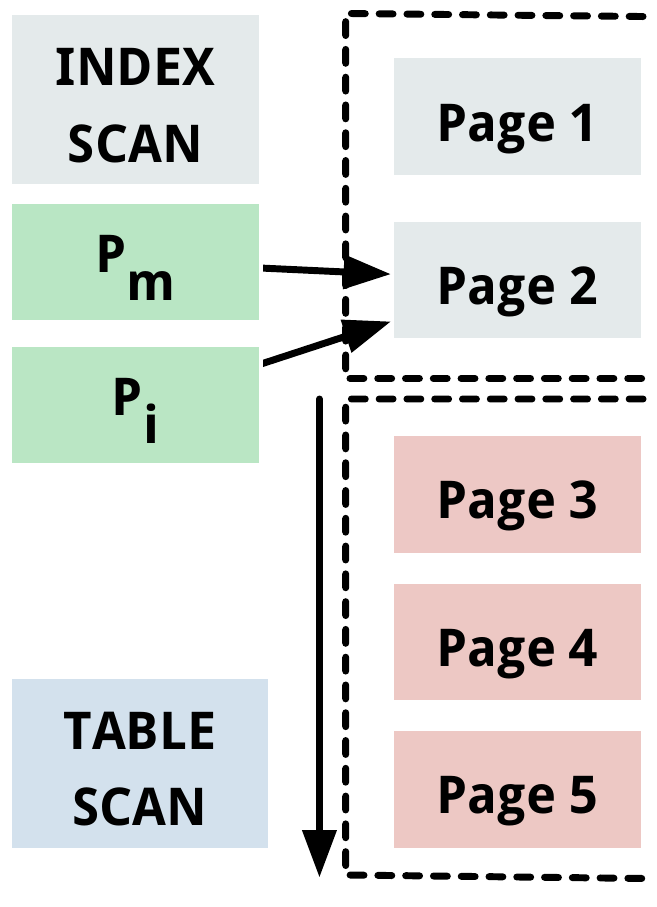}
        \label{fig:overlap-1}
    }
    \hfill
    \subfloat[$\rho_{m}>\rho_{i}$]{
        \includegraphics[width=0.3\columnwidth]
                        {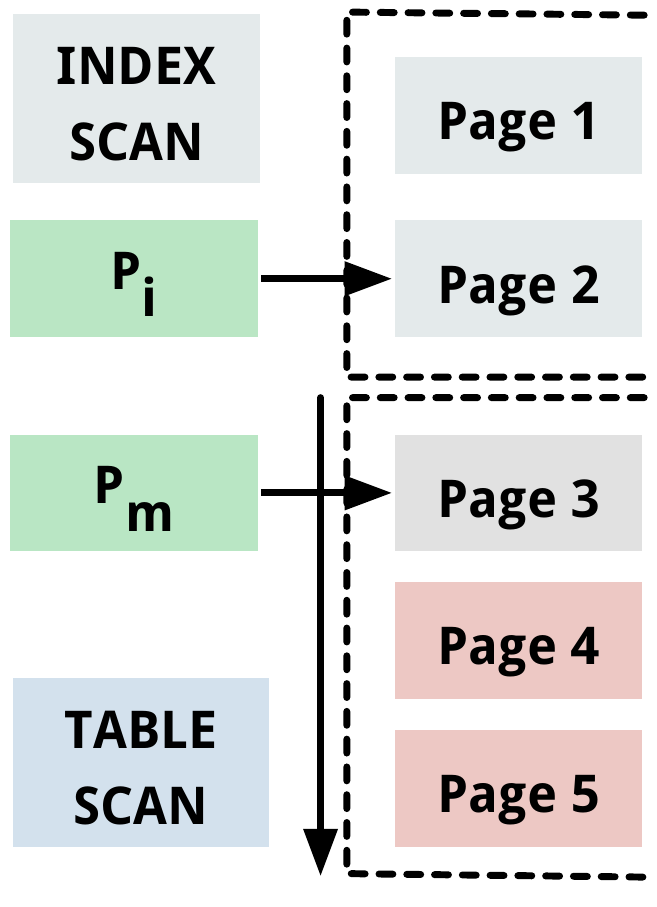}
        \label{fig:overlap-2}
    }
    \hfill
    \subfloat[$\rho_{m}<\rho_{i}$]{
        \includegraphics[width=0.3\columnwidth]
                        {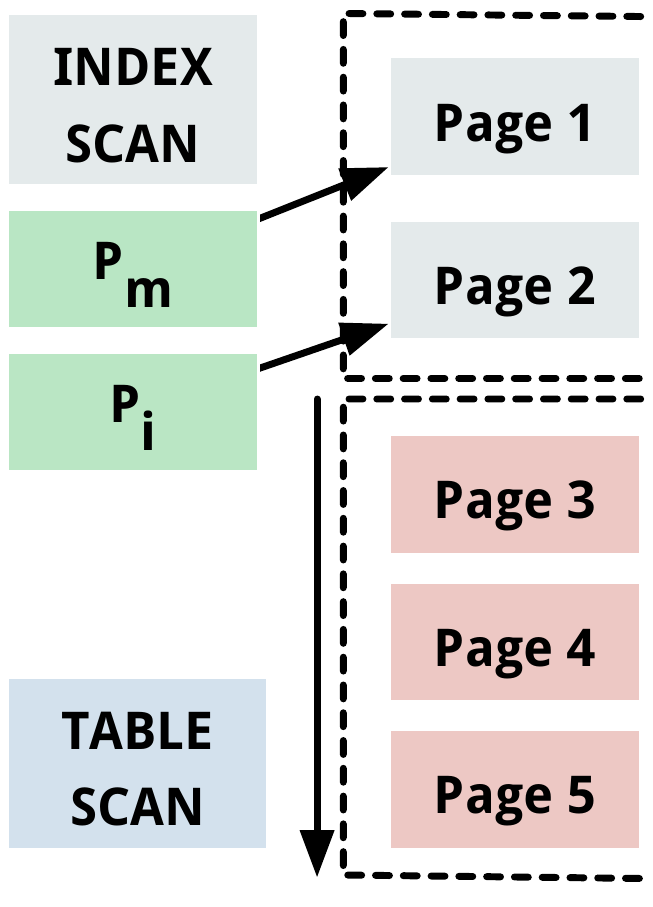}
        \label{fig:overlap-3}
    }
    \caption{
        \textbf{Hybrid Scan Operator} --
        \normalfont{Different scenarios depicting the \sysPage from which the hybrid scan
        operator must start scanning the table in order to ensure that it
        returns each matching tuple once and exactly once.}
    }
    \label{fig:hybrid-scan-operator}
\end{figure}

The DBMS assigns each page a unique \textit{page identifier} ($\rho$).
While populating an ad-hoc index on a table, the index tuner traverses its
\sysPage{s} in ascending order with respect to the page identifier.
During query processing, the hybrid scan operator goes over the \sysPage{s} in
the same order. While scanning the index, the operator keeps track of the
identifiers of two pages:
(1) the page with the largest identifier that contains a matching tuple
($\rho_{m}$) and (2) the page with the largest identifier that has already been
completely indexed ($\rho_{i}$).

When the index scan finishes, the operator determines the page from where it
should begin the table scan by computing $\max(\rho_{m}, \rho_{i}+1)$.
This is because, as shown in \cref{fig:hybrid-scan-operator}, 
there are only three possibilities: 
(1) $\rho_{m}$=$\rho_{i}$, 
(2) $\rho_{m}$>$\rho_{i}$, or 
(3) $\rho_{m}$<$\rho_{i}$.
In all of these, starting the table scan from $\max(\rho_{m},
\rho_{i}+1)$ ensures that it returns all matching tuples at least once.
Only when $\rho_{m}>\rho_{i}$, the operator additionally needs to remove
duplicate tuples present in the \textit{overlapping \sysPage}. The
\texttt{EMPLOYEE} table example illustrates this scenario.
After the index scan, the operator first populates the matching tuples contained
in the overlapping third \sysPage in a sorted data structure. 
Then, during the subsequent table scan over the same \sysPage, it checks for
duplicate tuples using the sorted data structure, and skips returning them twice.
\\ \vspace{-0.1in}

\textbf{Concurrency Control \& Updates:}
\peloton employs the multi-version concurrency control (MVCC) protocol for
scheduling transactions. 
The DBMS records the versioning meta-data alongside each tuple, and uses it to
determine whether a tuple version is visible to a transaction.
While handling updates, it appends the modified tuple versions in the table.
Since the tuner will subsequently index these tuples, the DBMS does not
propagate the changes to the ad-hoc indexes built by the tuner.
The hybrid scan operator, therefore, reads updates made by concurrent
transactions that are visible as per the MVCC protocol during its table
scan. This design allows the hybrid scan operator to work 
well on a wide variety of HTAP workloads, including those containing
long-running scan queries.
\\ \vspace{-0.1in}

\textbf{Query Optimization:}
We now describe how we extended the query optimizer to use hybrid scan.
\peloton's optimizer examines the query's predicates and the access paths
available on its referenced tables, and estimates a cost for each 
plan~\cite{selinger79,soliman14}. It uses a hybrid scan access path similar to
how it employs an index scan access path. It picks a hybrid scan only for
processing high selectivity queries, and switches to a table scan otherwise.

\subsection{Value-Agnostic \& Value-Based Scan}
\label{sec:hybrid::discussion}

The value-agnostic hybrid scan operator differs from the value-based operator
employed by prior indexing
approaches~\cite{petraki15,voigt13,idreos07}.
With the value-agnostic operator, the tuner can better control the 
index construction overhead. It incrementally populates the index by
adding entries only for a fixed number of \sysPage{s} at a time.
This decouples the indexing overhead from the value distributions of the
attributes being indexed.
The partially-built index only needs to keep track of the number
of \sysPage{s} that have already been fully indexed, so that the operator can
determine the \sysPage from which it must begin the table scan operation.
The main limitation of the value-agnostic operator is that the query processing
time drops linearly over time, unlike the logarithmic drop observed with the
value-based operator~\cite{idreos07,athanassoulis16}.

With the value-based operator, after the tuner completely populates a
sub-domain of an index, it can skip the table scan operation while
handling subsequent queries for the indexed sub-domain.
This shrinks the query processing time but causes latency spikes when the size of the
sub-domain accessed is large. As such, 
the tuner may not be able to react in time to future workload shifts.
Further, the index must maintain meta-data about the sub-domains that have
already been indexed, so that the operator can safely skip the table scan operation when
scanning indexed sub-domains. To prevent latency spikes, we 
therefore employ the value-agnostic operator. 
We note that both hybrid scan operators help shrink the \textit{adaptation
time} of the DBMS in comparison to the \full scheme employed by online indexing
approaches~\cite{valentin00,chaudhuri98b}. This is because they enable the DBMS
to immediately make use of partially-built indexes.

\section{Predictive Decision Logic}
\label{sec:model}

We next present how the tuner's predictive DL decreases its \textit{detection 
time}. The complexity of the index tuning problem can be
attributed to two factors~\cite{finkelstein88}.
First, the tuner must balance the performance gains of an index during
query processing against the overhead of maintaining it while executing queries 
that update the index. The tuner must, therefore, adapt the index configuration
based on the current workload mixture. On a read-intensive
workload, it should build additional indexes for accelerating query processing.
But when the workload shifts to write-intensive queries, it should drop a subset
of these indexes to reduce the index maintenance costs.

Second, the tuner must take the storage space consumed by the index into
consideration. This is an important constraint even when there are no updates.
Indexes typically consume a significant portion of the total memory footprint of
the DBMS, particularly on OLTP workloads.
This is because OLTP applications often maintain several indexes per table 
to ensure that the queries execute quickly. For instance, a previous study
found that indexes consume more than half of the storage space used by an 
in-memory DBMS for OLTP workloads~\cite{zhang16}. The tuner must, therefore,
ensure that the indexes fit within the available memory.

\begin{algorithm}[t!]
    \small
    \caption{\Pred Indexing Algorithm}
    \label{alg:index}
    \begin{algorithmic}
    \Require
    Recent queries $\mathcal{Q}$ ($\mathcal{R}$ and $\mathcal{W}$),
    Index configuration $\mathcal{I}$,
    Index storage budget $\mathcal{S}$
    
    \State \codeComment{Executed once during every tuning cycle}
    \Function{EVOLVE-INDEX-CONFIGURATION}{$\mathcal{Q}, \mathcal{I},
    \mathcal{S}$} 
    \State \codeComment{Stage I : Workload classification} 
    \State Label $\mathcal{L}$ = Workload-Classifier($\mathcal{Q}$)
    \State \codeComment{Stage II : Action Generation}
    \State Candidate indexes $\mathcal{C}$ = Candidate-Indexes($\mathcal{Q}$) 
    \State Overall utility $\mathcal{O}$ = Overall-Utility($\mathcal{I}$,
    $\mathcal{C}$, $\mathcal{Q}$)     
    \State Index configuration $\mathcal{I}'$ =
    Index-Knapsack($\mathcal{L}$, $\mathcal{S}$, $\mathcal{I}$, $\mathcal{C}$,
     $\mathcal{U}$, $\mathcal{O}$) 
    \State Apply lightweight changes to reach index
    configuration $\mathcal{I}'$ 
    \State \codeComment{Stage III : Index Utility Forecasting}
    \State Forecasted utility $\mathcal{U}$ =
    Holt-Winters-Forecaster($\mathcal{U}$, $\mathcal{O}$)
\EndFunction  
\end{algorithmic}
\end{algorithm}

The predictive machine-learning (ML) model contains three principal components: 
(1) workload classifier, 
(2) action generator, and 
(3) index utility forecaster.
These components perform the three steps that correspond to the
\textit{observe-react-learn} template that an agent typically 
follows in reinforcement-learning~\cite{kaelbling96}.
\cref{alg:index} presents an overview of the \pred indexing algorithm.

During every index tuning cycle, the classifier examines the recent query
workload that is tracked by a lightweight \textit{workload monitor}, 
and determines the type of the workload~\cite{bog11,levandoski13,das16}. 
Based on the workload classification, the recently executed queries,
the utility of the indexes present in the current index configuration, 
and the recent query workload, the action generator performs a set of
\textit{actions}, that involve building and dropping indexes.
These actions change the index configuration of the DBMS, and the utility of
this \textit{state transition} is fed back as input by the index utility forecaster
to the action generator as the reinforcement signal~\cite{kaelbling96}. 
Through this systematic trial-and-error, the tuner learns to choose actions that
increase the utility of the index configuration, 
thereby improving the performance of the DBMS. We next describe these three
model components in further detail.

\subsection{Workload Classification}
\label{sec:model::classifier}

The type of the DBMS's current query workload is an important consideration
for the index tuning process. This is because the benefits of an index
during the workload's \textit{read-intensive} phases could be outweighed 
by the cost of maintaining it during the \textit{write-intensive} phases. 
The tuner constructs an index only when it expects it to be beneficial for the
current and near-future workloads.

\begin{figure}[t!]
    \centering
    \includegraphics[width=0.8\columnwidth]{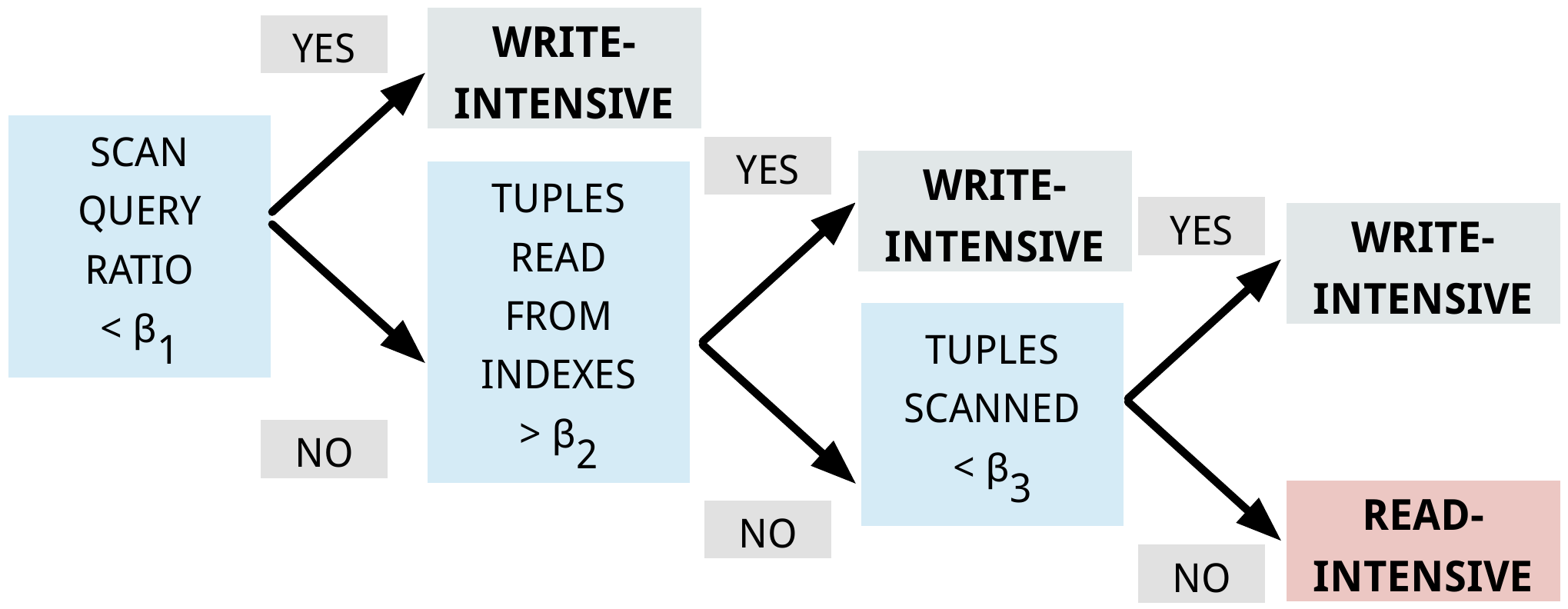}
    \caption{
        \textbf{Workload Classifier} --
        \normalfont{The pruned decision tree used by the
        tuner to determine the type of the query workload.}
    }
    \label{fig:decision-tree}
\end{figure}

We construct a workload classifier in the form of a \textit{decision tree} using the classification 
and regression tree (CART) algorithm~\cite{cart84}. This algorithm constructs a binary tree, where 
each
node contains a feature and threshold that yields the largest information gain.
We train the classifier using \textit{workload snapshots} collected by running a
set of OLTP and OLAP benchmarks in \peloton~\cite{difallah2013}.
Each snapshot contains meta-data about $k$ recently executed queries, 
and is manually assigned a \textit{classification label}.
The collected meta-data contains a set of domain-specific \textit{features},
such as ratio of the number of \texttt{SELECT} statements to that of \texttt{UPDATE},
\texttt{INSERT}, and \texttt{DELETE} statements~\cite{elnaffar02}.
We construct a training set of snapshots that are each assigned one of the 
following labels:

\squishitemize
    \item \textbf{Write-Intensive:} The
    workload is characterized by a large number of short on-line transactions that 
    mutate the database. The maintenance cost of an ad-hoc index might
    outweigh its benefits during query processing on this type of workload.
    \item \textbf{Read-Intensive:} The workload is
    characterized by a small number of complex read-only queries that often involve aggregations.
    The benefits of ad-hoc indexes is often greater than
    the index maintenance overhead on this type of workload.
\squishend

We use a decision tree for the classifier because it is better suited for this
classification task compared to other ML techniques by virtue of
its \textit{interpretability}; it is easier to understand the classification
rules and explain the classifier's decisions. 
Based on our domain-specific knowledge, we configure the DBMS's monitor
to collect the following features in every workload snapshot:

\squishitemize
    \item
    \textbf{Ratio of scan queries to mutators:} The ratio of the
    number of \texttt{SELECT} to \texttt{UPDATE}, \texttt{INSERT}, and \texttt{DELETE} 
    statements.
  
    \item
    \textbf{Ratio of tuples accessed using indexes:}
    The ratio of tuples that are accessed via indexes rather than 
    directly from the table. It tends to be higher for write-intensive   
    workloads.
  
    \item
    \textbf{Number of tuples scanned:} The average number of tuples 
    scanned (higher for read-intensive workloads).
\squishend

\cref{fig:decision-tree} presents the pruned decision tree for workload
classification. The higher-level nodes in the decision tree correspond to 
features that are more important in the classification process. 
The ratio of queries to mutators is the crucial feature for workload classification.
It is important to not take features that are dependent on the system
utilization and database configuration into consideration, such as metrics like system throughput 
and cache hit ratio.

The tuner uses the classifier during each tuning cycle to determine the current
query workload type. Based on this classification, it determines how to next evolve the index
configuration. If it detects a write-intensive workload, it prunes the
configuration by dropping ad-hoc indexes of low utility\footnote{The
tuner assigns high utility values to indexes that are used for processing 
\texttt{UPDATE} statements even in a write-intensive workload, and will thus
refrain from dropping them.}.
In contrast, if the classifier detects a read-intensive workload, then the tuner
expands the configuration by building auxiliary indexes to accelerate the analytical queries. 
The classifier only emits classifications that are supported
by a minimum threshold of query samples. This ensures that it is
robust during periods of low system utilization. We next describe how the tuner
generates actions for mutating the index configuration.

\subsection{Action Generation}
\label{sec:model::action-generator}

The action generator picks the actions that need to be taken to improve
the performance of the DBMS. These actions could either
involve building an index or dropping an existing index. 
We note that the following discussion is relevant for both \textit{primary} and
\textit{secondary} indexes. The generator picks these actions based on the
current workload classification, the recently executed queries, and the utility of the indexes
present in the current index configuration.
These actions mutate the index configuration of the DBMS, and the value of this
state transition is later estimated by the forecaster.

The lightweight workload monitor tracks the attributes that are accessed by each
query. The goal is to determine the set of attributes present in the statement
predicates and clauses that should be indexed by an ad-hoc index.
In particular, it keeps track of the attributes appearing in equal, 
range, and join predicates, the attributes present in \texttt{GROUP}~\texttt{BY} and
\texttt{ORDER}~\texttt{BY} clauses, and the remaining attributes referenced in the SQL
statement. The monitor stores this information for each individual table.
The generator then uses this information to enumerate various combinations of
these attribute sets, and determines the set of \textit{candidate indexes} 
that are not present in the current index configuration~\cite{valentin00}.
The set of candidate indexes includes both \textit{single-attribute} and
\textit{multi-attribute} indexes. We next describe our methodology for computing
the utility of an index.
\\ \vspace{-0.1in}

\textbf{Query Processing Utility:}
The tuner uses the query optimizer to obtain the estimated processing cost 
of a given query in the presence and absence of a candidate 
index~\cite{valentin00,chaudhuri98}.
We refer to the difference between the costs of executing a query in the
presence and absence of an index as its \textit{query processing utility} (QPU).
The tuner uses this information to determine whether it should construct a
particular candidate index or not.
QPU of a candidate index $\mathcal{I}$ is derived by evaluating its impact on
the set of scan queries $\mathcal{R}$ in the latest workload snapshot whose
underlying table scans can be accelerated using $\mathcal{I}$. 
Let us denote the cost of processing a query $r$ using only the existing
indexes as $\eta$($r$), and let $\eta$($r$, $\mathcal{I}$) represent the 
cost of processing the same query using $\mathcal{I}$ in addition to the
existing indexes ($\eta$($r$) $\geq$ $\eta$($r$, $\mathcal{I}$)). Then, QPU of
$\mathcal{I}$ is represented by:
$$\text{QPU}(\mathcal{I}, \mathcal{R}) = \sum\limits_{r \in
\mathcal{R}} (\eta(r) -\eta(r,I) )$$
\\ \vspace{-0.1in}

\textbf{Index Maintenance Cost:}
The tuner next considers the overhead of maintaining the candidate and
currently built indexes while executing statements that mutate 
the index. We refer to this overhead as the \textit{index maintenance cost}
(IMC). This is crucial for the DBMS to perform well on write-intensive
workloads. For each index, the tuner keeps track of the queries $\mathcal{W}$
that update, insert, or delete entries in the index.
Let us denote the cost of maintaining an index $\mathcal{I}$ while 
processing a query $w$ by $\tau$($w$, $\mathcal{I}$). 
The IMC of $\mathcal{I}$ is represented by:
$$\text{IMC}(\mathcal{I}, \mathcal{W}) = \sum\limits_{w \in
\mathcal{W}} \tau(w,\mathcal{I})$$

The \textit{overall utility} of an index is obtained by discounting the
index maintenance cost from the query processing utility of an index:
$$\text{Overall Utility}(\mathcal{I}, \mathcal{R}, \mathcal{W}) = 
\text{QPU}(\mathcal{I}, \mathcal{R}) - \text{IMC}(\mathcal{I}, \mathcal{W})$$

\textbf{Index Knapsack Problem:}
After determining the overall utility of the candidate and current indexes, as
shown in \cref{alg:index}, the tuner constructs a subset of the candidate
indexes and drops a subset of the currently built indexes, after 
taking their storage footprint into consideration. 
We can formulate it as a 0-1 {knapsack problem}~\cite{martello90}.
Let $\mathcal{C}$ represent the set of candidate and currently built
indexes. The set of candidate indexes includes: (1) indexes that are expected
to speed up queries in the near future by the index utility forecaster, and 
(2) indexes that accelerate newly encountered queries whose trends have not yet
been captured by the forecaster. For the latter subset, we bootstrap the 
utility of these indexes in the forecaster's model with their overall utility,
as illustrated in \cref{alg:index}. The tuner dampens the utility of redundant
indexes with correlated attributes by characterizing their
interactions~\cite{bruno07}.

By capturing the index utility patterns, the forecaster enables the tuner
to look ahead into the future and optimistically build indexes.
Let us assume that each index $c \in \mathcal{C}$ has a utility $U_{c}$ $\geq
0$ and an estimated storage footprint $S_{c}$ $\geq 0$. 
For newly suggested indexes, $U_{c}$ refers to the overall utility. 
In case of indexes that are either currently built or existed in the
past, it represents the forecasted utility. 
Let $B$ denote the storage space available
for storing indexes.
Then, the solution of the knapsack problem is a subset of 
indexes $\mathcal{C'} \subseteq \mathcal{C}$ with maximal utility:
$$\text{maximize}  \sum\limits_{c \in \mathcal{C'}} U_{c} \text{ such that}
\sum\limits_{c \in \mathcal{C'}} S_{c} \leq B $$


\textbf{Index Configuration Transition:}
Based on the classification by the workload classifier presented in
\cref{sec:model::classifier}, the tuner determines the minimum utility
($U_{min}$) required for an index to exist.
It scales up $U_{min}$ while handling a write-intensive workload, 
and scales it down on a read-intensive workload.
After determining the indexes to add and drop, the action generator
applies those changes in the index configuration. 
We refer to this step as a state transition towards the desired final
index configuration state.
The tuner amortizes the overhead of this state transition over a set of tuning
cycles to ensure that the DBMS's performance is predictable.

\subsection{Index Utility Forecasting}
\label{sec:model::forecaster}

\begin{figure}
    \centering
    \fbox{\includegraphics[width=0.6\columnwidth, trim=2.75cm 0.5cm 0cm 0.15cm]
                            {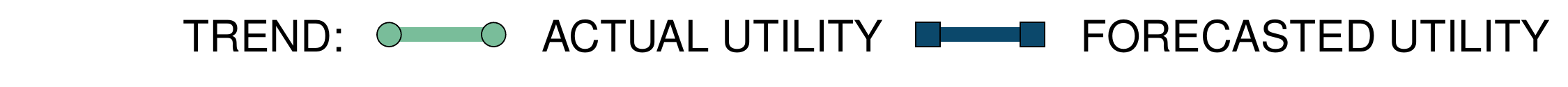}}
    \\[-0.25ex] 
    \includegraphics[width=0.48\textwidth]
                    {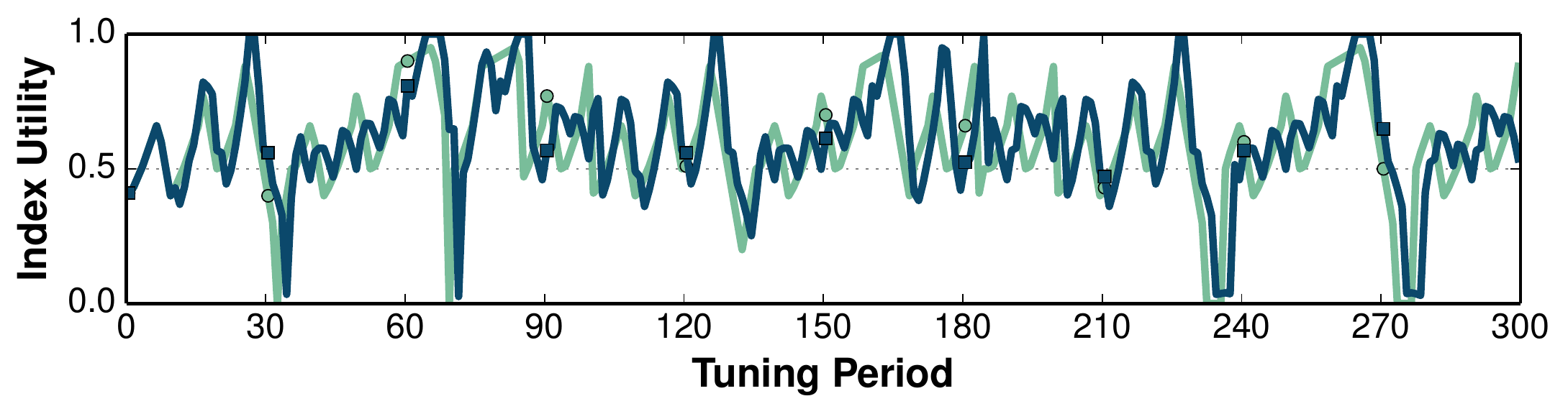}
    \caption{
        \textbf{Learning Utility} --
        \normalfont{Forecasting index utility using the
        Holt-Winters method with multiplicative seasonality.}
    }
    \label{fig:holt-winters}
\end{figure}

The final component of the tuner is the index utility forecaster which 
predicts the utility of indexes in the near future. 
As shown in \cref{alg:index}, the forecasted utility of the index configuration
after the state transition is fed back as input to the action generator during
the next tuning cycle. Using this feedback, the tuner learns to pick actions
that increase the long-run sum of the reinforcement signal, which is the
overall utility of the index configuration.
\\ \vspace{-0.1in}

\textbf{Holt-Winters Seasonal Method:}
To forecast the utility of the indexes, the tuner uses a variant of
\textit{exponential smoothing}.  
This method predicts the index utility by computing a weighted average 
of past utility observations with the weights decaying exponentially as the
observations get older.
The \textit{Holt-Winters seasonal method} extends the basic exponential
smoothing technique to allow forecasting of data with a trend besides capturing
seasonality~\cite{holt57,winters60}.
This method is comprised of the forecast equation along with three smoothing
equations. The three smoothing equations estimate the \textit{level} $l_{t}$, 
the \textit{trend} $b_{t}$, and the \textit{seasonal} component $s_{t}$ of a
time series.
\begin{align*}
\text{Forecast equation: } & \hat{y}_{t + h | t} = (l_{t} + h b_{t}) 
s_{t-m+h_{m}}
\\
\text{Level equation: } & l_{t} = \alpha (y_{t}/s_{t-m}) + (1 - \alpha) (l_{t-1}
+ b_{t-1} ) \\
\text{Trend equation: } & b_{t} = \beta (l_{t} - l_{t-1}) + (1 - \beta) 
b_{t-1} \\
\text{Seasonal equation: } & s_{t} = \gamma (y_{t}/(l_{t-1} + b_{t-1})) + (1 -
\gamma) s_{t-m}
\end{align*}

Here, $\alpha$, $\beta$, and $\gamma$ are the smoothing parameters for the
utility level, trend, and seasonality respectively, and $\alpha$, $\beta$,
$\gamma$ $\in$ $[0, 1]$. $m$ denotes the period of the seasonality.

Using the forecast equation, the tuner predicts the utility of indexes in 
the near future after factoring in the trend and seasonality aspects of the
time series, as shown in \cref{fig:holt-winters}. 
The accuracy of the forecaster's predictions with respect to an index increases
over time as it obtains more observations.
The Holt-Winters method dampens the utility of indexes that are not beneficial
in the recent past. Over time, the tuner drops the less useful indexes to make
space for building new indexes that are more useful during query
processing. Even after dropping an index, the tuner's forecaster retains the
model meta-data associated with that index. This allows it to predict that
index's utility in future when it is requested by the action generator.



We next describe the design of the
benchmark that we use for evaluating \pred indexing.

\section{\benchTuner Benchmark}
\label{sec:benchmark}

We developed a new open-source benchmarking suite to evaluate the efficacy of
\pred indexing with respect to other online indexing
approaches. The queries in this benchmark are derived from a HTAP workload in a
bus tracking mobile application~\cite{peloton}.
The \benchTuner benchmark extends prior physical design tuning benchmarking
suites proposed in \cite{olma17,alagiannis14,schnaitter09}
in three ways. It examines the impact of the index tuner under both regular and heavy
system loads, when the tuner throttles its tuning frequency to cope with
workload spikes.
Second, it evaluates the impact of automatically adapting both the storage
layout and the index configuration of the database in tandem, 
thus focusing on two key components of the physical design problem. 
Lastly, it characterizes the \pred index tuner's ability to leverage idle system
resources.

The \benchTuner database contains two tables: a \narrow table and a \wide
table. Each table contains tuples with a timestamp ($a_{0}$) and $p$ integer
attributes ($a_{1},\ldots,a_{p}$), each 4 bytes in size.
The \narrow table has $p$ = 20 attributes, and the \wide table has $p$ = 200
attributes.
Each attribute $a_{k}$ is an integer value from a Zipf distribution
in the range $[$1, 1m$]$. 
The benchmark can configure the selectivity of the predicates in a query while
constructing multi-attribute \textit{range} and \textit{equality predicates}
present in the mobile application's HTAP workload.
We next describe our methodology for constructing these queries that filter and
aggregate the dataset based on different attribute sets.

\subsection{Query Generation}
\label{sec:benchmark::query}
 
We construct workloads using two kinds of queries: (1) scan queries and (2)
update queries.
\\ \vspace{-0.1in}

\textbf{Scan Queries:}
We use three \texttt{SELECT} templates to construct queries with different
levels of workload complexity:

\squishitemize
    \item
    \textbf{Low-Complexity Scan (LOW-S):}
    It computes aggregates over a single table after applying a \textit{comparison predicate} 
    defined on a single attribute. This query template represents the simplest scenario for an index
    tuner.
	\vspace{1mm}\\    
    \begin{boxedminipage}{0.93\columnwidth}
    \texttt{\textbf{SELECT} $a_{1}$,$a_{2} + a_{3}$,$\ldots$,$\texttt{SUM}(a_{k})$
    \textbf{FROM} R\\
    \hspace*{4pt}\textbf{WHERE} $a_{i}\geq\delta_{1}$ \textbf{AND} $a_{i}\leq\delta_{2}$}
    \end{boxedminipage}
    \\
    
    \item
    \textbf{Moderate-Complexity Scan (MOD-S):}
    It is similar to the previous template except that the comparison predicate is now defined on a 
    combination of attributes. The tuner must build multi-attribute indexes to
    accelerate such queries.
    \vspace{1mm}\\    
    \begin{boxedminipage}{0.93\columnwidth}
    \texttt{\textbf{SELECT} $a_{1}$,$a_{2} + a_{3}$,$\ldots$, $\texttt{SUM}(a_{k})$
    \textbf{FROM} R\\ 
    \hspace*{4pt}\textbf{WHERE} $a_{i}\geq\delta_{1}$ \textbf{AND} $a_{i}\leq\delta_{2}$
    \textbf{AND} $a_{j}\geq\delta_{3}$ 
    \textbf{AND} $a_{j}\leq\delta_{4}$} 
    \end{boxedminipage}
    \\

    \item
    \textbf{High-Complexity Scan (HIGH-S):}
    Unlike the previous query templates, this query contains an equi-join operation over two 
    relations in addition to the multi-attribute comparison predicates. The
    query would, therefore, benefit from the presence of indexes defined on the attributes involved in the \textit{join predicate}.
    \vspace{1mm}\\    
    \begin{boxedminipage}{0.93\columnwidth}
    \texttt{\textbf{SELECT} X$.a_{1}$,$\ldots$,X$.a_{k}$, Y$.a_{1}$,$\ldots$,Y$.a_{k}$ \\
    \hspace*{4pt}\hspace*{4pt}\textbf{FROM} X, Y \\
    \hspace*{6pt}\textbf{WHERE} X.$a_{i}\geq\delta_{1}$ \textbf{AND}
    X.$a_{i}\leq\delta_{2}$ \textbf{AND} \\
    \hspace*{4pt}\hspace*{4pt}\hspace*{4pt}\textbf{AND} Y.$a_{j}\leq\delta_{4}$ \textbf{AND} 
    X.$a_{l}$ = Y.$a_{m}$} 
    \end{boxedminipage}
    \\

\squishend

We vary the selectivity and projectivity of the predicates in the scan queries
by altering $\delta_{1}$, $\delta_{2}$, and $k$.
\\ \vspace{-0.1in}

\textbf{Update Queries:}
We use three different query templates to construct updates and inserts
that correspond to different levels of workload complexity. These include:

\squishitemize
    \item
    \textbf{Low-Complexity Update (LOW-U):}
    It updates a random subset of attributes for tuples that satisfy a comparison predicate 
    defined on a single attribute. 
    \vspace{1mm}\\    
    \begin{boxedminipage}{0.93\columnwidth}
    \texttt{\textbf{UPDATE} R \textbf{SET}
    $a_{1}$=$v_{1}$,$a_{2}$=$v_{2}$,$\ldots$,$a_{k}$=$v_{k} + 1$ \\
    \hspace*{4pt}\textbf{WHERE} $a_{i}\geq\delta_{1}$ \textbf{AND} $a_{i}\leq\delta_{2}$}
    \end{boxedminipage}
    \\

    \item
    \textbf{High-Complexity Update (HIGH-U):}
    It is similar to the low-complexity update query except that the comparison predicate is now 
    defined on a combination of attributes.
    \vspace{1mm}\\        
    \begin{boxedminipage}{0.93\columnwidth}
    \texttt{\textbf{UPDATE} R \textbf{SET}
    $a_{1}$=$v_{1}$,$a_{2}$=$v_{2}$,$\ldots$,$a_{k}$=$a_{k} + 1$ \\
    \hspace*{4pt}\textbf{WHERE} $a_{i}\geq\delta_{1}$ \textbf{AND} $a_{i}\leq\delta_{2}$
    \textbf{AND} $a_{j}\geq\delta_{3}$ \textbf{AND}
    $a_{j}\leq\delta_{4}$} 
    \end{boxedminipage}
    \\

    \item
    \textbf{Insert (INS):}
    It inserts a set of tuples into the table. Unlike the \texttt{UPDATE} statement that involves 
    both table scan and modifications, \texttt{INSERT} statement only performs table modifications.
    \vspace{1mm}\\        
    \begin{boxedminipage}{0.93\columnwidth}
    \texttt{\textbf{INSERT} \textbf{INTO} R \textbf{VALUES} ($a_{0},
    a_{1},\ldots,a_{p})$ }
    \end{boxedminipage}
\squishend

\subsection{Workload Generation}
\label{sec:benchmark::workload}

We use the above queries to construct a set of workloads that exercise
different aspects of \pred indexing.
A workload is a sequence of queries with changing properties.
To clearly delineate the impact of the index configuration on the different
queries, we divide a sequence into \textit{phases} consisting of a fixed
number of queries that each correspond to a particular query type. 
This means that the DBMS executes the same query type (with different input
parameters) in one phase, and then switches to another one in 
the next phase.
\\ \vspace{-0.1in}

\textbf{Tuning Frequency:}
The benchmark examines the tuner's impact when it only operates during certain
time periods. The tuner usually runs frequently, and is likely to
react quicker to short-lived workload shifts. In contrast, when the load is
high, it runs less frequently and is likely to treat these ephemeral shifts
as noise, and demonstrates gains only on longer phases.
The benchmark examines the tuner's impact when it operates under four different
frequencies: \tunerF, \tunerM, \tunerS, and \tunerD. With the \tunerF, \tunerM,
and \tunerS configurations, the tuner runs on average once every 100, 1000, and
10000~ms. Index tuning is disabled with \tunerD, and serves as a baseline.
\\ \vspace{-0.1in}

\textbf{Shifting Workloads:}
We construct shifting workloads of different phase lengths to examine how
swiftly can the tuner adapt to these shifts.
If the total number of queries in a workload is $t$, and the phase length
is $l$, then the workload is comprised of $\frac{t}{l}$ phases.
The performance impact of tuning depends on the average phase length of the
workload. Stable workloads with longer phases benefit more from tuning. 
This is because after the tuner observes a set of queries based on a
query template, it builds an index to accelerate query processing. 
Subsequent queries of the same template would also be accelerated.
\\ \vspace{-0.1in}

\textbf{Hybrid Workloads:}
To evaluate the ability of the tuner to handle both scan and update queries,
we construct hybrid workloads wherein the tuner must factor in both 
the query processing utility and the maintenance cost of indexes. 
We construct four types of workload mixtures that vary the operations that the
DBMS executes. These mixtures represent different ratios of scans and updates:

\squishitemize
    \item \textbf{Read-Only:} 100\% \textit{scans}
    \item \textbf{Read-Heavy:} 90\% \textit{scans}, 10\% \textit{updates}
    \item \textbf{Balanced:} 50\% \textit{scans}, 50\% \textit{updates}
    \item \textbf{Write-Heavy:} 10\% \textit{scans}, 90\% \textit{updates}
\squishend

On the read-intensive mixtures, the DBMS benefits from the construction 
of ad-hoc indexes that accelerate query processing. The benefits are not as
pronounced on the write-intensive mixtures due to index maintenance costs. 
We construct the updates using the \texttt{UPDATE} query templates, and
configure the scans to be low-complexity queries.

\section{Experimental Evaluation}
\label{sec:exps}

We now present an analysis of our \pred indexing approach using
\peloton~\cite{peloton}.
\peloton is an in-memory MVCC DBMS that supports HTAP workloads. 
We integrated our index tuner as a background thread that 
runs periodically depending on the tuning frequency.
It decides both when and how to adapt the index configuration.
During every tuning cycle, it classifies the workload, generates and applies
physical design changes, and learns the utility of these changes. 
We extended the system's execution engine and query optimizer to
support the value-agnostic hybrid scan operator.
To perform a comparative evaluation, we integrated different  
indexing approaches in \peloton, including 
(1) online~\cite{bruno07,schnaitter06}, 
(2) adaptive~\cite{idreos07}, 
(3) self-managing~\cite{voigt13}, and 
(4) holistic~\cite{petraki15}.

We deployed \peloton on a dual-socket Intel Xeon E5-4620 server running Ubuntu
14.04 (64-bit). Each socket contains eight 2.6~GHz cores. It has 128~GB of DRAM and 20~MB of L3 
cache. For each experiment, we execute the workload
five times and report an average of the metrics. All transactions are 
executed with the default snapshot isolation level. 
We disable logging to ensure that our measurements only reflect query processing time.
Each table in the \benchTuner database contains 10m tuples, 
and the total size of the database is 8.8~GB.

We begin with an analysis of the ability of the tuner to recognize workload
trends using its forecaster.
We then investigate the benefits of using hybrid scan to leverage partially
built indexes. Next, we compare our tuner against the holistic indexing approach
on a HTAP workload.
We then demonstrate how the index tuner works seamlessly with the storage layout
tuner in \peloton. Lastly, we present an analysis of the tuner's ability
to adapt to shifting workloads. 

\subsection{Decision Logic}
\label{sec:exps::decision}

\begin{figure}[t!]
    \centering
    \fbox{\includegraphics[width=0.45\textwidth, trim=2.25cm 0.5cm 0cm 0.15cm]
                        {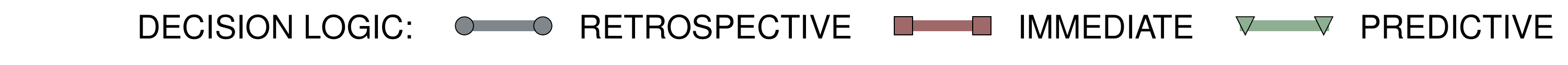}}
    \\[-0.2ex] 
    \includegraphics[width=0.48\textwidth]
                    {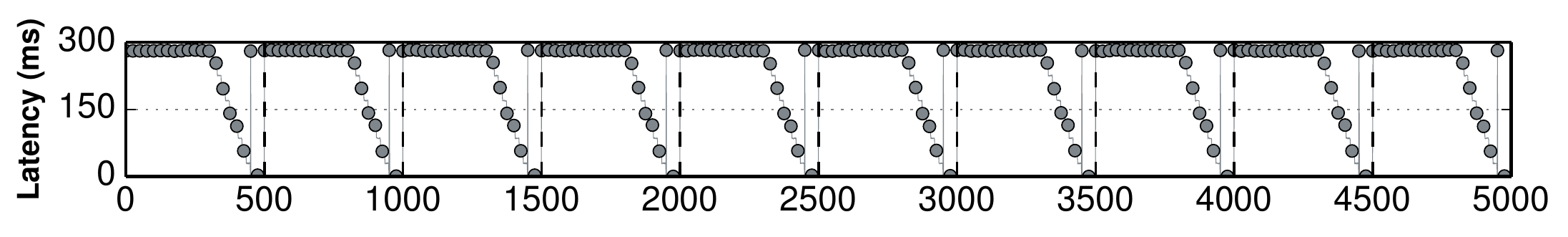}
    \hfill    
    \\[-1ex] 
    \includegraphics[width=0.48\textwidth]
                    {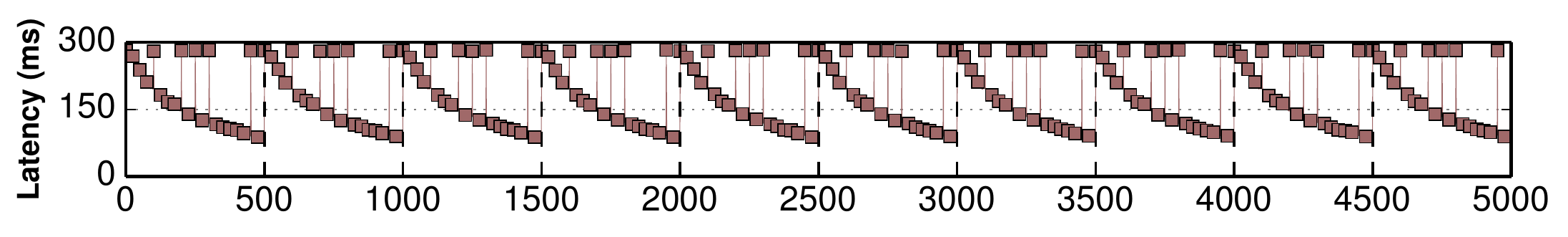}
    \hfill    
    \\[-1ex] 
    \includegraphics[width=0.48\textwidth]
                    {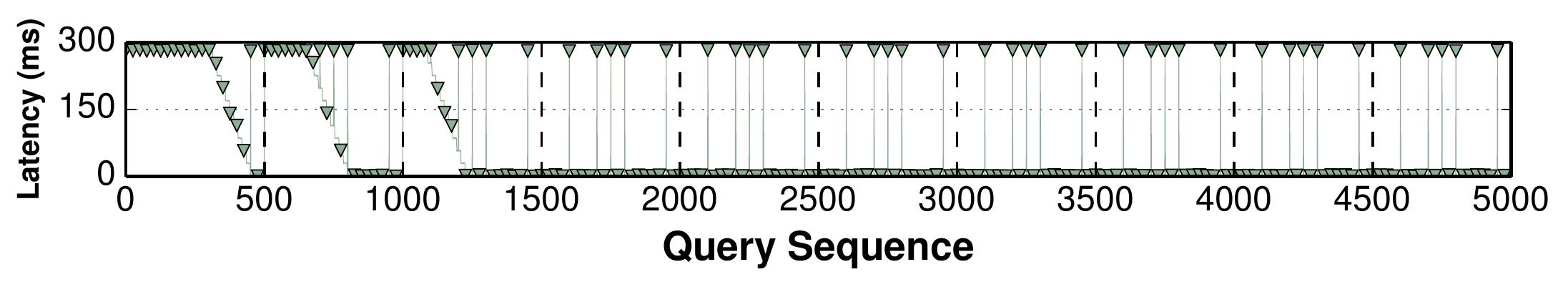}
    \caption{
        \textbf{Decision Logic} --
        \normalfont{Reaction times of the index tuner when using different types of decision
        logic for a HTAP workload.}
    }
    \label{fig:forecasting}
\end{figure}

We use a HTAP workload to examine the ability of the tuner to
recognize workload trends and adapt the index configuration ahead of time.
This workload contains 5000 moderate complexity scan queries with a phase length
of 500 queries. 
We throttle the client request throughput at the beginning of each phase to 
evaluate the tuner's ability to leverage idle system resources while it employs 
different types of decision logic: (1) retrospective DL, (2) predictive DL, and
(3) immediate DL.
We note that the index tuner must construct multi-attribute indexes to
accelerate the processing of these scan queries.
Most of the queries contain the same multi-attribute predicate, and will thus
benefit from the same index. A small fraction (1\%) of the queries are noisy
queries containing other predicates. 
Ideally, the tuner should recognize that it is not useful to 
build indexes to accelerate such queries.  We drop all the built indexes at the
end of each phase. This is meant to model a diurnal workload where indexes have
to be rebuilt every morning. 

\begin{figure*}[t!]
    \centering
    \fbox{\includegraphics[width=0.35\textwidth, trim=2.25cm 0.5cm 0cm 0.15cm]
                        {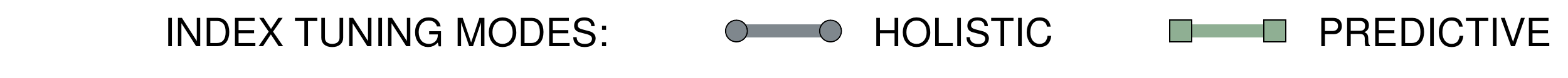}}
    \\[-0.3ex] 
    \includegraphics[width=0.95\textwidth]
                    {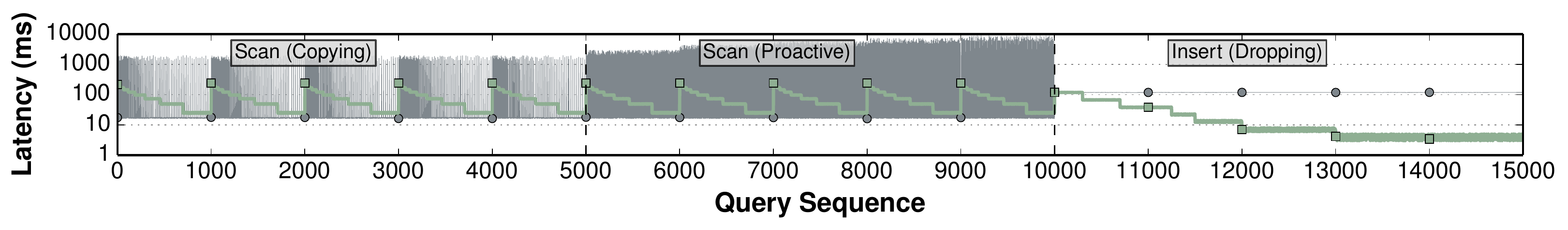}
    \caption{
        \textbf{Holistic Indexing} --
        \normalfont{Comparison of the performance impact of the holistic and the \pred
        indexing approaches on a HTAP workload.}
    }
    \label{fig:holistic}
\end{figure*}

The key observation from \cref{fig:forecasting} is that with \pred DL, the
tuner's forecaster captures the query pattern after observing 1500 queries.
It then makes use of idle system resources at the beginning of each phase to
build the index ahead of time, thus allowing the DBMS to start using the
index earlier within the phase.
With retrospective DL, the tuner only examines the recent $k$ queries in the
same phase, and does not proactively build indexes. 
This increases the tuner's detection time, and it finishes index construction
only later in the phase. Retrospective DL still guards against noisy queries.
In contrast, with immediate DL, the tuner constructs indexes to accelerate these
noisy queries, thereby slowing down the overall index construction speed and
increasing the adaptation time.
The cumulative time taken by the DBMS to execute this workload with \pred DL is
5.2$\times$ and 3.5$\times$ shorter than that taken with the retrospective and 
immediate schemes, respectively. This experiment shows that \pred DL can
both shrink the detection time of the tuner and guard against noisy queries.

\subsection{Hybrid Scan}
\label{sec:exps::hybrid}

We next examine the performance impact of different approaches for leveraging a
partially-built index: \full, \vap, and \vbp. We consider three
workloads with varying \textit{affinity levels} with respect to the index
sub-domains accessed by the queries: (1) very high, (2) high, 
and (3) moderate levels. Queries in workloads with higher affinity levels
target a smaller number of sub-domains.
The workloads contain 5000 moderate complexity scan queries with a phase length
of 500 queries. We configure the number of different sub-domains accessed by the
queries to be 2, 5, and 10 to create workloads with varying affinity
levels.
We construct a variant of \vbp that prevents latency spikes by decoupling 
index construction from query processing. With this scheme, the tuner
incrementally populates entries associated with a sub-domain over several tuning
cycles, instead of immediately adding all the entries. 

\begin{figure}[t!]
    \centering
    \fbox{\includegraphics[width=0.45\textwidth, trim=2.5cm 0.5cm 0cm 0.15cm]
                            {images/legend/legend_motivation.pdf}}
    \\[-0.2ex] 
    \includegraphics[width=0.48\textwidth]
                    {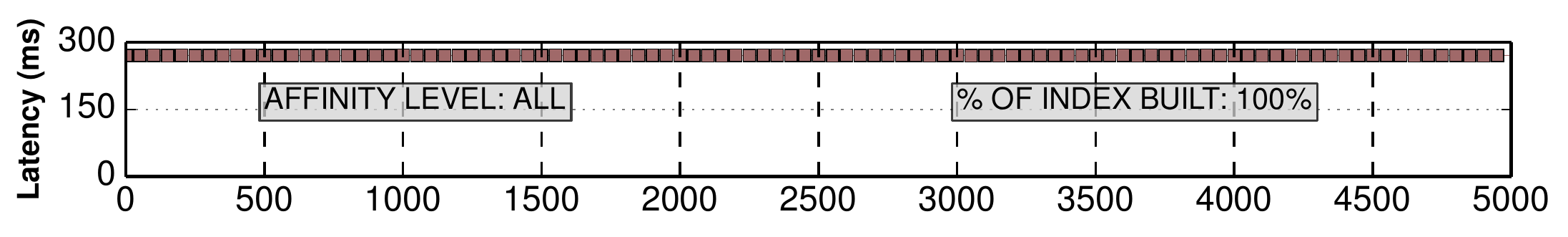}
    \hfill            
    \\[-1ex] 
    \includegraphics[width=0.48\textwidth]
                    {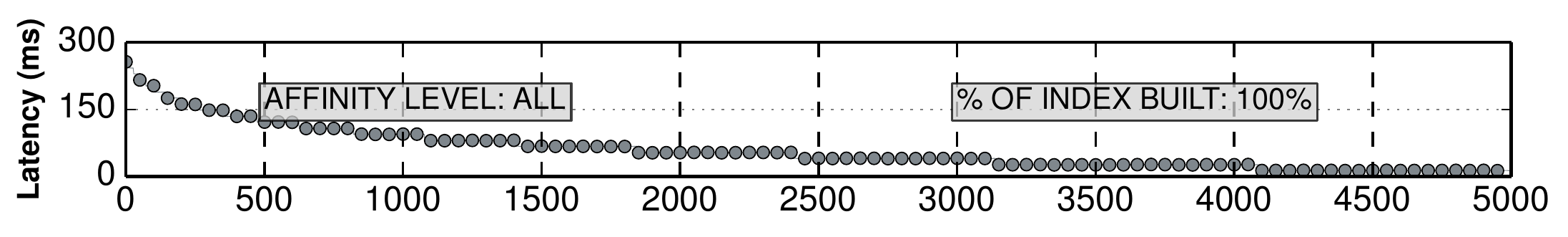}
    \hfill                    
    \\[-1ex] 
    \includegraphics[width=0.48\textwidth]
                    {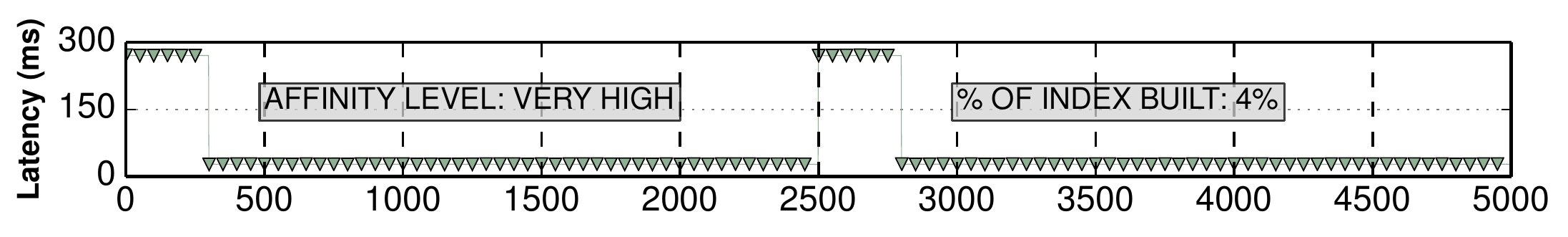}
    \hfill                    
    \\[-1ex] 
    \includegraphics[width=0.48\textwidth]
                    {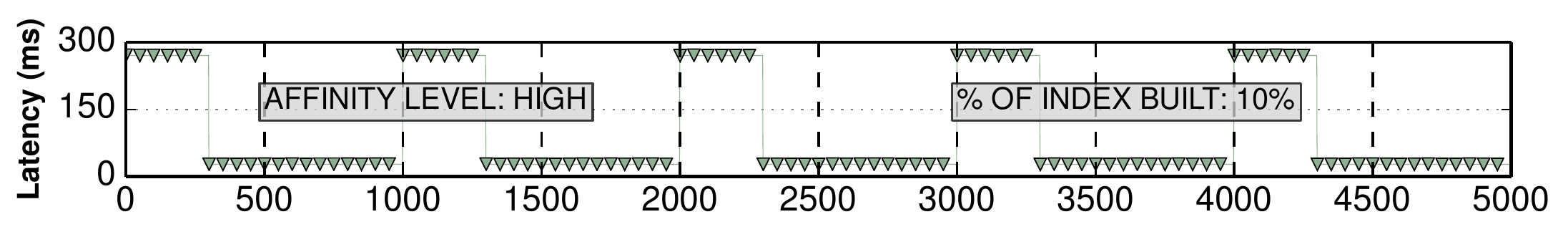}
    \hfill                    
    \\[-1ex] 
    \includegraphics[width=0.48\textwidth]
                    {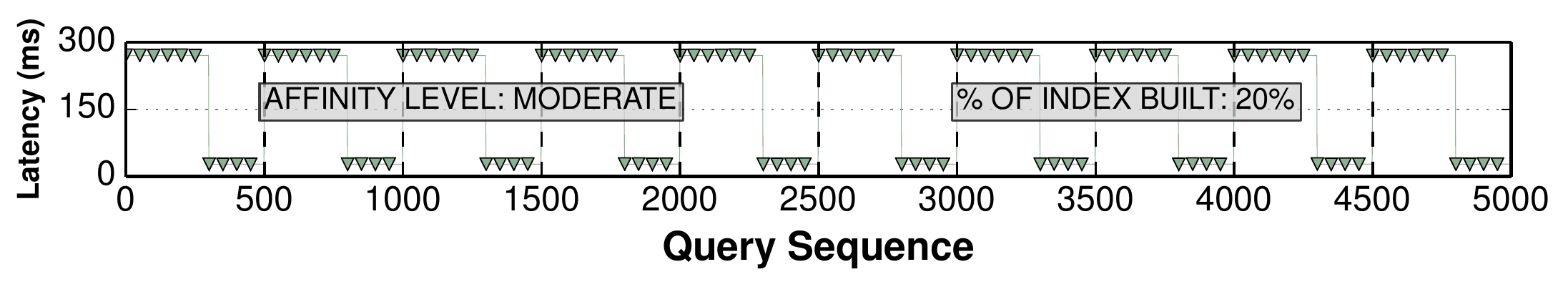}
    \hfill
    \\[-0.9ex] 
    \caption{
        \textbf{Hybrid Scan Operators} --
        \normalfont{Comparison of the performance impact of the different approaches for
        leveraging partially-built indexes. 
        }
    }
    \label{fig:hybrid-comparison}        
\end{figure}

The results in \cref{fig:hybrid-comparison} show that unlike \vbp, 
the behavior of \vap and \full is unchanged on workloads with different
affinity levels.
This is because with \vbp, the DBMS can leverage the partially-built
index only when the sub-domain accessed by the query has been already indexed.
But \vap enables the DBMS to make use of the partially-built index irrespective
of the sub-domain accessed by the query.

The cumulative time taken to execute the workloads with moderate and high
affinity levels with \vap is 3.1$\times$ and 1.7$\times$ shorter than that taken
with \vbp, respectively. This is because, unlike \vbp, \vap can accelerate query
processing even on workloads with lower affinity levels. 
On the very high affinity level workload, \vap tuner takes 1.05$\times$ longer
than that taken with \vbp. This is because the \vbp tuner populates the
frequently accessed sub-domains faster than \vap.
We observe that with \vbp and \full, the index is fully populated near
the end of the workload. In contrast, with \vap, only 4\%, 10\%, and 20\% of the
index is built depending on the number of different sub-domains accessed
in the workload. The key observation from this experiment is
that \vap allows the DBMS to leverage the index even when the sub-domain 
is not yet completely indexed.

\subsection{Holistic Indexing}
\label{sec:exps::holistic}

We now demonstrate how \pred indexing improves the performance of the DBMS in
comparison to holistic indexing.
Holistic indexing is an always-on \vbp scheme where the tuner optimistically
chooses which indexes to build next and then
populates them when the DBMS is idle~\cite{petraki15}.
Although \pred indexing shares many of the goals of
holistic indexing, it differs in three ways. It employs a value-agnostic hybrid
scan operator instead of a value-based one. It adopts a \pred DL instead of an
immediate DL. Lastly, it decouples index construction from query processing.
For this experiment, we implemented holistic indexing with a random index
selection strategy in \peloton and compare it against our approach~\cite{petraki15}.
We run a workload consisting of three \textit{segments}, each of which
contains 5000 queries based on multiple query templates. 
The first two segments consist of scan queries of moderate complexity, while the
last one contains insert queries.

The results in \cref{fig:holistic} show that on the first segment 
with holistic indexing, there are latency spikes that are as high as 
4$\times$ the latency of a table scan. 
When the tuner encounters a query accessing a sub-domain that has not yet been
indexed, it immediately starts indexing it while processing the
query~\cite{idreos07}.
Although this approach accelerates the execution of subsequent queries with the
same template, it involves heavyweight physical design changes. In contrast, we
do not observe such spikes with \pred indexing, since it amortizes the
index construction overhead across several tuning cycles.

To leverage idle system resources, holistic indexing advocates a proactive
approach towards building indexes even on attributes that have not
been queried yet~\cite{petraki15}. This can, however, increase the overall index
construction overhead, and thereby cause latency spikes during subsequent query
processing. This is illustrated during the second scan segment in
\cref{fig:holistic}. \Pred DL mitigates this problem by populating indexes on
attributes only after observing several queries that access them.

Lastly, on the third segment comprising of insert queries, the \pred tuner's
classifier detects the workload shift, and its action generator periodically
drops indexes of limited utility over time. This shrinks the insert query's
latency, as the DBMS needs to update fewer indexes. The holistic 
tuner does not drop any indexes since (by design) it drops them only when they
exceed its storage budget. The cumulative time taken to execute this
workload with \pred indexing is 7.7$\times$ shorter than that taken with
holistic indexing.

\begin{figure}[t!]
    \centering
    \fbox{\includegraphics[width=0.9\columnwidth, trim=2.25cm 0.5cm 0cm 0.15cm]
                        {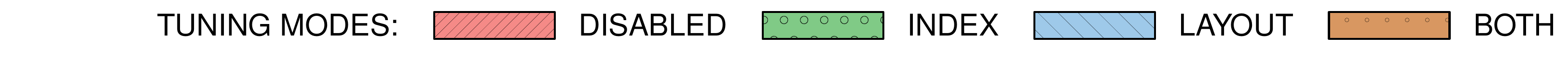}}
    \\[-0.25ex] 
    \includegraphics[width=0.95\columnwidth]
                    {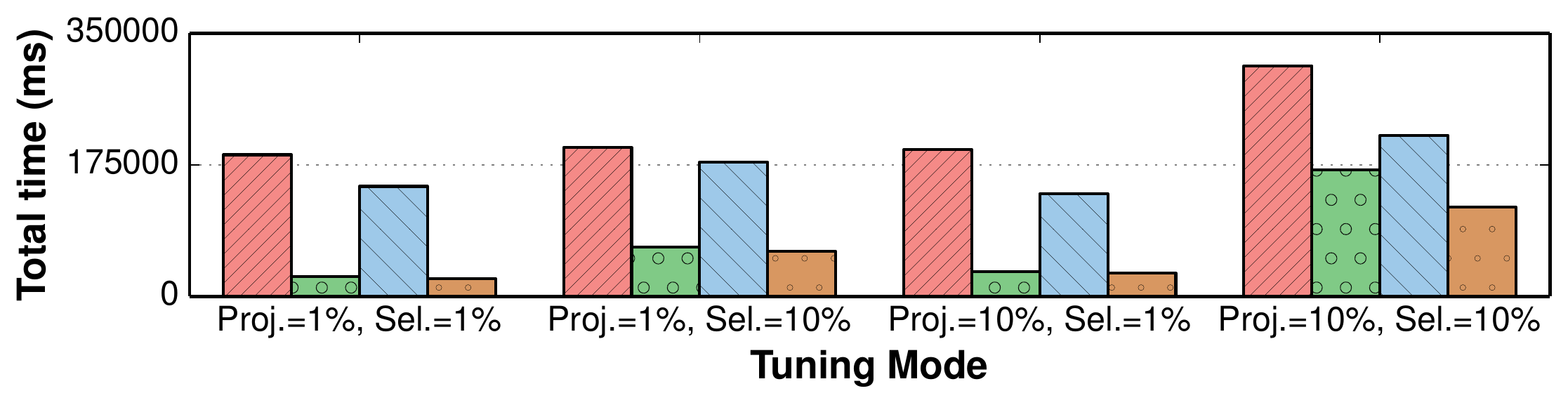}
    \caption{
        \textbf{Storage Layout \& Index Tuning} --
        \normalfont{The impact of index tuning in tandem with storage layout tuning on the query 
        processing performance.}
    }
    \label{fig:layout}
\end{figure}

\subsection{Storage Layout \& Index Tuning}
\label{sec:exps::layout}

This experiment examines the ability of the index tuner to work with
other physical design tuners.
We investigate the impact of the index tuner when used in tandem with the
storage layout tuner. This layout tuner accelerates query execution by
co-locating the attributes accessed together in a query in
memory~\cite{alagiannis14,arulraj16}.
This enables the DBMS to make better use of memory bandwidth by fetching only
the relevant attributes during query processing. 
We evaluate four tuning modes to measure the impact of index and layout tuning
on the system's performance:

\squishitemize
    \item \textbf{Disabled:} Both index and layout tuning are disabled.
    \item \textbf{Index:} Only index tuning is enabled.
    \item \textbf{Layout:} Only layout tuning is enabled.
    \item \textbf{Both:} Both index and layout tuning are enabled.
\squishend

We consider the behavior of the layout and index tuners under different
projectivity and selectivity settings on a read-only workload comprising of
moderate-complexity scan queries on the \wide table.
We vary both the projectivity and selectivity of scan query from 1\% to 10\%.
We measure the total time taken by the DBMS to execute the workload, including
the time spent on tuning the layout and index configuration.

The results in \cref{fig:layout} show that the index and layout tuners together are able to 
reduce the workload's
execution time more than that what they accomplish independently.
Under high projectivity and selectivity settings, we observe that while index
and layout tuning independently speed up query execution by 1.9$\times$ and
1.5$\times$, respectively, they reduce query execution time by 2.7$\times$ when
used in tandem. The layout tuner incrementally morphs the table to a hybrid
storage that collocates the query's projection attributes.
On average, the layout tuner takes 2.6~ms to transform the
layout of a \sysPage containing 1000 tuples. Concurrently, the index tuner
incrementally builds indexes to quickly retrieve the matching
tuples. During one iteration of \cref{alg:index}, the index tuner takes on average 5.2~ms to 
populate a set of indexes with entries associated with a
\sysPage.

The impact of the tuners is more prominent under low projectivity and
selectivity settings, as shown in \cref{fig:layout}. In this case, 
index and layout tuning together shrink the execution time by 7.8$\times$.
We attribute this to the higher gains from index and layout tuning when 1\%
of the attributes and 1\% of the tuples are respectively projected and selected
from the table. This highlights the importance of continuously improving the
physical design of the DBMS using many small steps. 

\subsection{Tuner Adaptability}
\label{sec:exps::adaptability}

We next analyze how the tuner adapts to shifting HTAP workloads.
We examine the impact of the tuner when operating under four different tuning
frequencies: \tunerF, \tunerM, \tunerS, and \tunerD. We consider two workload
mixtures, each comprising of 5000 queries, and we vary the phase length of the
workload from 50 to 500 queries.

\begin{figure}[t!]
    \centering    
    \fbox{\includegraphics[width=0.45\textwidth, trim=2.75cm 0.5cm 0cm 0.15cm]
                            {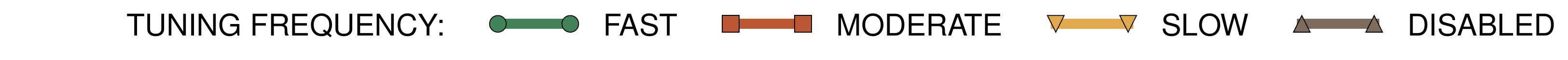}}
    \\[-1ex] 
    \subfloat[Read Only]{
        \includegraphics[width=0.22\textwidth]
                        {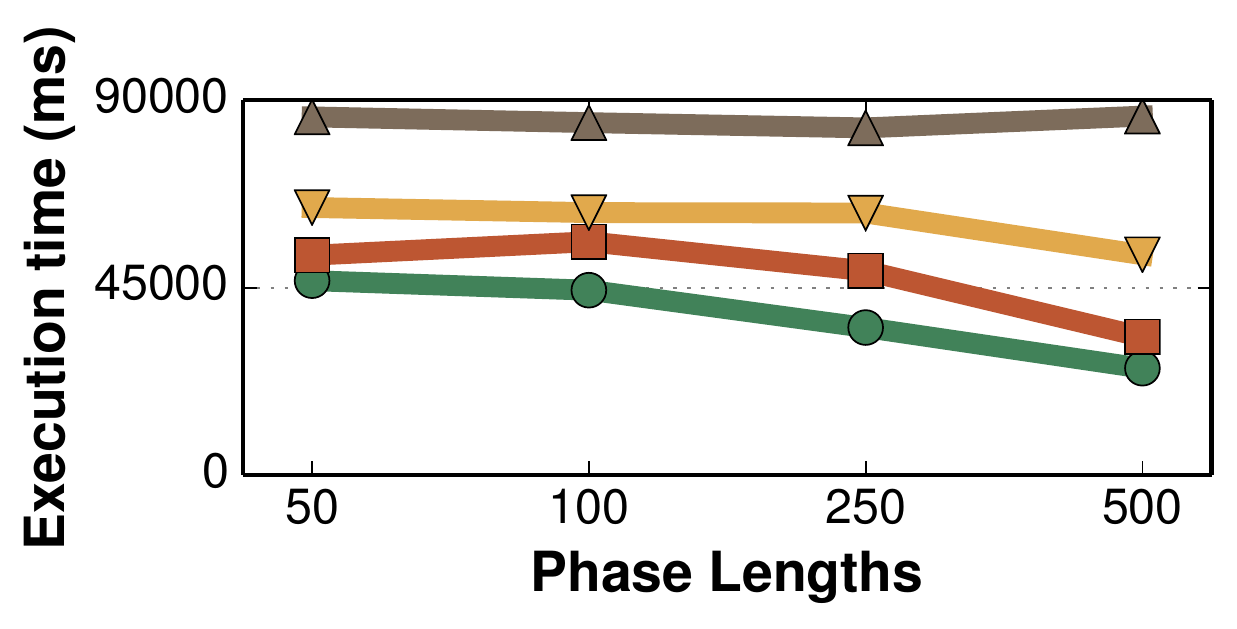}
        \label{fig:reflex-simple-read-only}
    }
    \hfill
    \subfloat[Write Heavy]{
        \includegraphics[width=0.22\textwidth]
                        {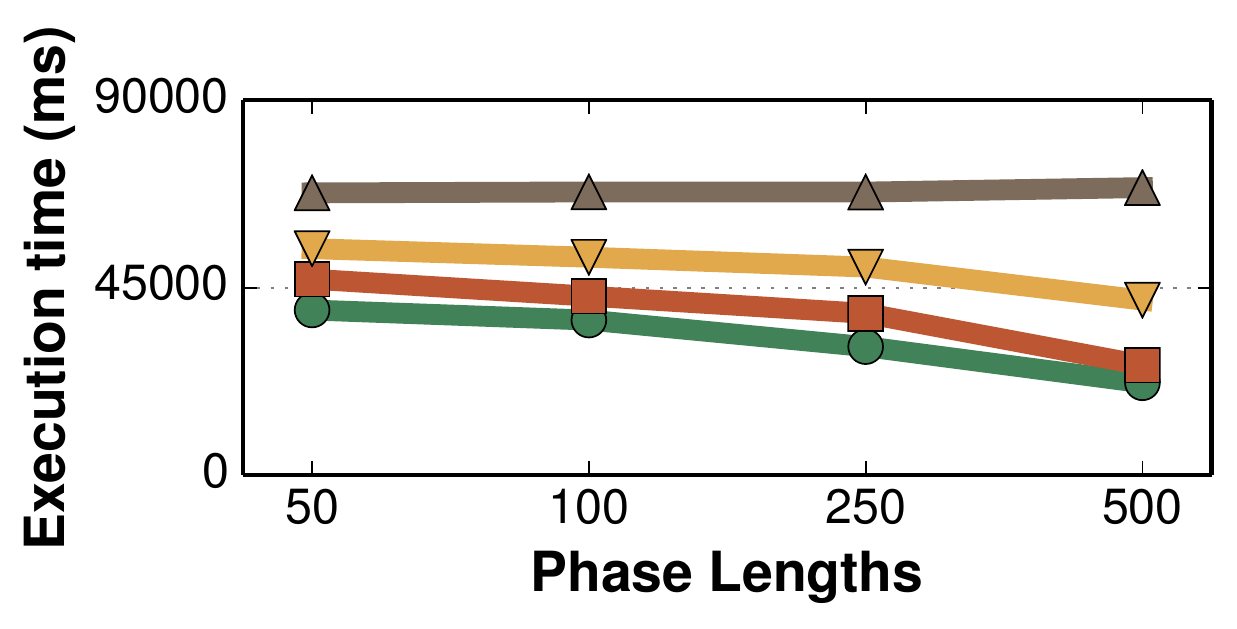}
        \label{fig:reflex-simple-write-heavy}
    }    
    \caption{
        \textbf{Tuner Adaptability} --
        \normalfont{The impact of the index tuning on the query processing time under
        different phase length settings. The execution engine runs different
        workload mixtures, each comprising of scan queries of varying
        complexity, on the \narrow table.}
    }
    \label{fig:adaptability}
\end{figure}


\cref{fig:reflex-simple-read-only,fig:reflex-simple-write-heavy} show the
results for the read-only and write-heavy workloads under different phase
length settings. The most notable observation is that the benefits are greater on 
workloads with longer phases. When the phase length is 500
queries, \tunerF outperforms \tunerD by 3.4$\times$. This is because after the
tuner constructs the appropriate index in a phase, all the subsequent
queries within the same phase benefit from the index.
We observe that \tunerM and \tunerS outperform \tunerD by 2.6$\times$ and
1.6$\times$, respectively. The impact of the tuner is less prominent under these
configurations because it takes more time to build the indexes due to 
lower tuning frequencies.

\section{Related Work}
\label{sec:related::work}

The design of \pred indexing benefited from prior work on online index tuners,
offline index advisors, self-managing indexing, and holistic indexing.
\\ \vspace{-0.1in}

\textbf{Online Index Tuners:}
\label{sec:related::online}
The problem of automatically tuning the index configuration of a database to
improve its performance on evolving query workloads has been studied for 
several decades~\cite{hammer76}. 
More recently, Bruno and Chaudhuri present an online algorithm that dampens
physical design oscillations and takes index interactions into
consideration~\cite{bruno07}.
COLT is an online index-tuning framework that dynamically
lowers its overhead when the system has converged to an optimal index
configuration~\cite{schnaitter06}.
All of this work present tuners that leverage the indexes during query
processing only after fully populating them. In contrast, \peloton uses 
the index with the hybrid scan operator even before it is fully
built, thereby improving its reaction time.
\\ \vspace{-0.1in}

\textbf{Self-Managing Indexing:}
\label{sec:related::smix}
Self-managing indexing (SMIX) is a \vbp scheme that can dynamically expand and
shrink indexes based on the workload~\cite{voigt13}. Unlike adaptive
indexing that incrementally converges to a full index, 
SMIX also shrinks the index by dropping less frequently accessed 
entries. However, SMIX only refines indexes during query processing and adopts
an immediate DL. This can cause latency spikes and increases the reaction
time. Unlike \pred indexing, it does not support range queries.
\vspace{-0.1in}

\section{Conclusion}
\label{sec:conclusion}

This paper presented \pred indexing that uses reinforcement learning 
to predict the utility of indexes, and continuously refines the index
configuration of the database to handle evolving HTAP workloads.
We proposed a lightweight value-agnostic hybrid scan operator that allows the
DBMS to leverage partially-built indexes during query processing. 
Our evaluation showed that the \pred index tuner learns over time to choose
physical design changes that increase the overall utility of the index
configuration. 
We demonstrated that the index and storage layout tuners in \peloton
work in tandem to incrementally optimize two key components of the database's
physical design without requiring any manual tuning.
Our evaluation showed that \pred indexing improves the throughput of \peloton on
HTAP workloads by 3.5--5.2$\times$ compared to other state-of-the-art indexing
approaches.

\balance

\bibliographystyle{IEEEtran}
{\small
\raggedright
\bibliography{predictive}
}


\end{document}